\documentclass[]{revtex4}
\usepackage{hetstyle}

\begin{document}
\title{Modes of wave-chaotic dielectric resonators}

\author{H.~E. Tureci}
\affiliation{Department of Applied Physics, Yale University, New
Haven, CT 06520, USA}

\author{H.~G.~L. Schwefel}
\affiliation{Department of Applied Physics, Yale University, New
Haven, CT 06520, USA}

\author{Ph.~Jacquod}
\affiliation{D\'epartement de Physique Th\'eorique,
Universit\'e de Gen\`eve, CH-1211 Gen\`eve 4, Switzerland}

\author{A.~Douglas Stone}
\affiliation{Department of Applied Physics, Yale University, New
Haven, CT 06520, USA}

\date{\today}
\begin{abstract}
Dielectric optical micro-resonators and micro-lasers represent a realization of a wave-chaotic system, where the lack of symmetry in the resonator shape leads to non-integrable ray dynamics. Modes of such resonators display a rich spatial structure, and cannot be classified through mode indices which would require additional constants of motion in the ray dynamics. Understanding and controlling the emission properties of such resonators requires the investigation of the correspondence between classical phase space structures of the ray motion inside the resonator and resonant solutions of the wave equations. We first discuss the breakdown of the conventional eikonal approximation in the short wavelength limit, and motivate the use of phase-space ray tracing and phase space distributions. Next, we introduce an efficient numerical method to calculate the quasi-bound modes of dielectric resonators, which requires only two diagonalizations per N states, where N is approximately equal to the number of half-wavelengths along the perimeter. The relationship between classical phase space structures and modes is displayed via the Husimi projection technique. Observables related to the emission pattern of the resonator are calculated with high efficiency.
\end{abstract}


\maketitle
\section{Introduction}
A promising approach to making compact and high-Q optical resonators is to base them on ``totally internally reflected" modes of dielectric micro-structures. Such devices have received considerable attention as versatile components for integrated optics and for low threshold micron-scale semiconductor lasers\cite{chang_book,yamamoto93}. The interest in such resonators for applications and for fundamental optical physics has motivated the extension of optical resonator theory to describe such systems.

All optical resonators are open systems described by modes characterized by both a central frequency and a width (their ratio giving the Q-factor of the mode). In a mirror-based resonator the set of resonant frequencies is determined by an optical path-length for one round-trip along a path determined by the mirrors within the resonator; the width is determined by the reflectivity of the mirrors, diffraction at the mirror edges and by absorption loss within the resonator.  Accurate analytic formulas can be found for the resonator frequencies and for the electric field distribution of each mode using the methods of Gaussian optics\cite{siegman_book}. The modes are characterized by one longitudinal and two transverse mode indices (in three dimensions).  These mode indices play the same role mathematically for the electromagnetic wave equation as the good quantum numbers play in characterizing solutions of the wave equation of quantum mechanics.

For fabricating optical resonators on the micron scale, using total internal reflection from a dielectric interface for optical confinement is convenient as it simplifies the process. Such dielectric resonators define no specific optical path length; many different and potentially non-closed ray trajectories can be confined within the resonator.  An important point, emphasized in the current work, is that in such resonators there typically exist many narrow resonances characterized by their frequency and width, but such resonances often cannot be characterized by any further modes indices.  This is the analog of a quantum system in which there are no good quantum numbers except for the energy. We shall see that the way to determine if a given mode has additional mode indices (other than the frequency), is to determine whether it corresponds to regular or chaotic ray motion.  We will present below an efficient numerical method for calculating all of the resonances of a large class of dielectric resonators; we will also describe the surface of section and Husimi-Wigner projection method to determine the ray dynamics corresponding to such a mode.

Although both DBR-based and edge-emitting optical resonators rely on reflectivity from a dielectric interface (at normal incidence), we will use the term dielectric resonator (DR) to refer to resonators that rely on the high reflectivity of dielectric bodies to radiation incident from within the dielectric near the critical angle for total internal reflection. This is the only class of resonators we will treat below. We immediately point out that totally-internally-reflected solutions of the wave equation only exist for infinite flat dielectric interfaces; any curvature or finite extent of the dielectric will allow evanescent leakage of propagating radiation from the optically more dense to the less dense medium.  As a dielectric resonator is a finite dielectric body embedded in air (or in a lower index medium) it will of necessity allow some evanescent leakage of all modes, even those which from ray analysis appear to be totally-internally reflected.

A very large range of shapes for DRs have been studied during the recent years. By far the most widely studied are rotationally symmetric structures such as spheres, cylinders and disks. The reason for this is that the wave equation is separable and the solutions can be written in terms of special functions carrying three modes indices. The narrow (long-lived) resonances correspond to ray trajectories circling around the symmetry axis near the boundary with angle of incidence above total internal reflection; these solutions are often referred to as ``whispering gallery" (WG) modes or morphology-dependent resonances. In this case, due to the separability of the problem, it is straightforward to evaluate the violation of total internal reflection, which may be interpreted as the tunneling of waves through the angular momentum barrier\cite{johnson93,noeckel_thesis}. Micron-scale, high-Q micro-lasers were fabricated in mid-80s and early 90s based on such cylindrical \cite{mccall92,slusher93,levi92} (disk-shaped) ($Q \sim 10^4-10^5$) and spherical\cite{CollotLBRH93} ($Q \sim 10^8-10^{12}$) dielectric resonators. However, the very high Q value makes these resonators unsuitable for micro-laser applications, because such lasers invariably provide low-output power and furthermore, unless additional guiding elements are used, the lasing output is emitted isotropically.

As early as 1994\cite{NockelSC94} one of the current authors proposed to study dielectric resonators based on smooth deformations of cylinders or spheres which were referred to as ``asymmetric resonant cavities" (ARCs). The idea was to attempt to combine the high Q provided by near total internal reflection with a breaking of rotational symmetry leading to directional emission and improved output coupling.  General principles of non-linear dynamics applied to the ray motion (to be reviewed below) suggested that there would be only a gradual degradation of the high-Q modes, and one might be able to obtain directional emission from deformed whispering gallery modes. Experimental\cite{NockelSCGC96,Mekis95,ChangCSN00} and theoretical\cite{nature} work since that initial suggestion has confirmed this idea, although the important modes are not always of the whispering gallery type\cite{science,gornik1,gmachl1,rex02,sblee1}.

Calculating the modal properties of deformed cylindrical and spherical resonators presents a much more challenging theoretical problem. Unless the boundary of the resonator corresponds to a constant coordinate surface of some orthogonal coordinate system, the resulting partial differential equation will not be solvable by separation of variables. The only relevant separable case is an exactly elliptical deformation of the boundary, which turns out to be unrepresentative of generic smooth deformations.  Using perturbation theory to evaluate the new modes based on those of the cylindrical or spherical case is also impractical, as for interesting deformations and typical resonator dimensions (tens of microns or larger) the effect of the deformation is too large for the modes of interest to be treated by perturbation theory. The small parameter in the problem for attempting approximate solutions is the ratio of the wavelength to the perimeter $\lambda/2 \pi R= (kR)^{-1}$. Eikonal methods\cite{kravtsov_book} and Gaussian optical methods\cite{siegman_book} both rely on the short wavelength limit to find approximate solutions.  The Gaussian optical method can be used to find a subset of the solutions for generic ARCs, those associated with stable periodic ray orbits, as explained in detail in Ref.\cite{Tureci02}. The eikonal method can also be used to find a subset of the modes of ARCs, if one has a good approximate expression for a local constant of motion; an example of this is the adiabatic approximation used by N\"ockel and Stone\cite{nature}.  However a large fraction of the modes in ARCs are not describable by either of these methods.  The breakdown of the Gaussian optical methods is easily seen as a fully chaotic system will have only unstable periodic orbits and the solutions one obtains by the Gaussian method near unstable periodic ray orbits are inconsistent\cite{Tureci02}. The failure of eikonal methods is more subtle and really arises from the possibility of chaotic ray motion in a finite fraction of the phase space.  Often optics textbooks and even standard research references treat the eikonal method as being of completely general applicability; we therefore will devote the next section of this paper to an explanation of the failure of eikonal methods for resonators with arbitrary smooth boundaries. In section~(\ref{sect_refbilliard}) we describe the phase space methods which indicate that this failure is generic.

In section~(\ref{sect_resprob}) we present the formulation of the resonance problem and in section~(\ref{sect_wavereduct}) the reduction of the Maxwell's equations to the Helmholtz Equation for the resonators we study. The failure of all standard short wavelength approximation methods to describe the solution of wave equations in finite domains with arbitrary smooth boundaries has led to the problem of ``quantizing chaos" in the context of the Schr\"odinger equation and the Helmholtz Equation (although our problem is somewhat different due to the dielectric boundary conditions on this equation). Although substantial progress has been made using periodic orbit methods in obtaining approximations for the density of states of fully chaotic systems, these methods do not yield individual solutions of the wave equation.  It is therefore of great importance in this field to develop efficient numerical methods which can be used to calculate and interpret the resonance properties. In section~(\ref{sect_scatquant1}) we present a highly efficient numerical method for ARCs, adapted from the S-matrix methods developed in the field of quantum chaos.  In section~(\ref{sect_rootsearch}) we display a range of resonant solutions for a partially chaotic dielectric resonator and in section~(\ref{sect_husimi}) we show how to perform the Husimi projection of the real-space numerical solutions so obtained into phase space in order to interpret them in terms of ray dynamics. The calculation of experimental observable relating to emission patterns from micro-lasers are discussed in section~(\ref{sect_FF}).  Finally, in section~(\ref{sect_qbm}), we show examples of the main types of modes one encounters in wave-chaotic dielectric resonators generally and specifically in ARCs.

\section{Failure of eikonal methods for generic dielectric resonators}
\label{sect_torus_quant}
The use of classical ray theory to describe monochromatic, high-frequency  solutions of the wave equation is described in various references\cite{keller_book,babic_book,kravtsov_book}. The connection between rays and waves is standardly derived in the context of the Helmholtz equation
\be
\left(\bm{\nabla}^2 + n^2(\bm{x})k^2 \right)\psi(\bm{x}) = 0;
\label{hheqintro}
\ee
the wave equation for the resonator problem will be reduced to this
equation in section~(\ref{sect_wavereduct}).  The eikonal approach uses the asymptotic
ansatz
\be
\psi(\bm{x}) \sim \ex{ikS(\bm{x})}\sum_{\nu=0}^{\infty}
\frac{A_{\nu}(\bm{x})}{k^{\nu}}
\label{asymptotic1}
\ee
in the limit $k \ra \infty$. Inserting Eq.~(\ref{asymptotic1}) into
Eq.~(\ref{hheqintro}) one finds to lowest order in the asymptotic
parameter $1/k$, the eikonal equation
\be
(\nabla S) ^2 = n^2(\bm{x})
\label{eqeikonal}
\ee
and the transport equation
\be
2 \bm{\nabla} S \cdot \bm{\nabla} A_0  + A_0 \nabla^2 S = 0
\label{eqtransp}
\ee

Note at this point we only assume one eikonal $S(\bm{x})$ and one
amplitude $A(\bm{x})$ at each order in the expansion. We will also
specialize to a uniform medium of dielectric constant $n$.  In this
framework, each wave solution $\psi(\bm{x})$ corresponds to a {\em family
of rays} defined by the vector field
\be
\bm{p}(\bm{x}) = \bm{\nabla} S(\bm{x})
\ee
where the field has a fixed magnitude, $ |\nabla S | = n$. The solution for the function $S (\bm{x})$ can be found by the specification of initial value boundary conditions on an open curve $\mcal{C}:\bm{x}=\bm{x}(s)$  and propagating the curve using the eikonal equation. Such an initial value solution can thus be extended until it encounters a point at which two or more distinct rays of the wavefront converge; at or nearby such a point will occur a focus or caustic at which the amplitude $A$ will diverge and in the neighborhood of which the asymptotic representation becomes ill-defined. (A caustic is a curve to which all the rays of a wavefront are tangent; if the curve degenerates to a point it is a focus\cite{kravtsov_book}).  This causes only a local breakdown of the method and can be handled by a number of methods. At a distance much greater than a wavelength away from the caustic the solutions are still a good approximation to the true solution of the initial value problem.

In contrast, to find asymptotic solutions on a bounded domain $D$ with boundary value conditions, one must introduce more than one eikonal at each order in the asymptotic expansion.  We will illustrate the important points here with Dirichlet boundary conditions on the boundary $\partial D$. However the basic argument holds for any linear homogeneous boundary conditions and, with minor modifications, for the matching conditions relevant for uniform dielectric resonators of index of refraction $n$ with boundary shape $\partial D$, within an infinite medium of index $n=1$. For the discussion of Dirichlet boundary conditions we will set the index $n=1$ for convenience within the domain $D$.  The leading order in the asymptotic expansion of the solution takes the form
\be
\psi(\bm{x}) = \sum_{m=1}^{N} A_m(\bm{x}) \ex{ikS_m(\bm{x})}
\label{introuscansatz}
\ee
with $N \geq 2$.  It is easily checked that there must be more than one term (eikonal) in the solution in order to have a non-trivial solution; if there were only one term in the expression for $\psi (\bm{x})$ then any solution which vanished on the boundary would vanish identically in $D$ due to the form of the transport equation.

The question we now address is the following. For what boundary shapes $\partial D$ in two dimensions do there exist approximate solutions of the form Eq.~(\ref{introuscansatz}) which are valid everywhere in $D$ except in the neighborhood of caustics (which are a set of measure zero)?

First we note that with Dirichlet boundary conditions we have a hermitian eigenvalue problem and so we know that solutions will only exist at a discrete set of real wavevectors $k$.  In the eikonal theory the quantization condition for $k$ arises from the requirement of single-valuedness of $\psi (\bm{x})$ and will be reviewed briefly below.  Here our primary goal is to show that the existence of eikonal solutions to the boundary value problem is intimately tied to the nature of the ray dynamics within the region $D$.  Moreover for the case of fully chaotic ray dynamics this connection shows that eikonal solutions {\em do not exist}. We will prove this latter statement by showing a contradiction follows from assuming the existence of eikonal solutions in the chaotic case. This argument will be a ``physicist's proof" without excessive attention to full mathematical rigor.

The proposed solution for $\psi (\bm{x})$ posits the existence of $N$ scalar functions $S_m(\bm{x})$ each of which satisfy the eikonal equation, $(\nabla S_m(\bm{x}))^2 = 1$ and which, while themselves not single-valued on the domain $D$, allow the construction of single-valued functions $\psi (\bm{x})$ and $\nabla \psi (\bm{x})$.  Moreover, for the asymptotic expansion to be well-defined, the ``rapid variation" in $\psi (\bm{x})$ must come from the largeness of $k$; i.e.\ to define a meaningful asymptotic expansion in which terms are balanced at each order in $k$ the functions $S_n$ cannot vary too rapidly in space. From the eikonal equation itself we know that $|\nabla S_m| = 1$, but we must also have that the curvature $\nabla^2 S_m \ll k$ for the asymptotic solution to be accurate.  This condition fails within a wavelength of a caustic, as one can check explicitly, e.g.\ for the case of a circular domain $D$; but for a solvable case like the circle it holds everywhere else in $D$.

It is convenient for our current argument to focus on $\nabla \psi$, instead of $\psi $ itself. Consider an arbitrary point $x_0$ in $D$ where $\nabla \psi (\bm{x}_0) \neq 0$; to leading order in $k$ and away from caustics
\be
\nabla \psi (\bm{x}_0)= ik \sum_m^N A_m(\bm{x}_0) \nabla S_m
(\bm{x}_0) e^{ikS_m(\bm{x}_0)}.
\label{gradpsi}
\ee
The $N$ unit vectors $\nabla S_m (\bm{x}_0) \equiv \hat{p}_m$ define $N$ directions at $x_0$ which are the directions of rays passing through $x_0$ in the stationary solution. An important point is that due to the condition on the curvature just noted, these directions are constant at least within a neighborhood of linear dimension $ \lambda = 2\pi/k$ around $x_0$. Choose one of the ray directions, call it $\hat{p}_1 $ and follow the gradient field $\nabla S_1 $ to the boundary $D$.  For a medium of uniform index (as we have assumed) the vector $\nabla S_1 $ is strictly constant in both direction and magnitude along a ray.  Thus one can find the direction of $\nabla S_1 $ at the boundary and calculate its ``angle of incidence'', $ \hat{n} \cdot \nabla S_1$, where $\hat{n}$ is the normal to the boundary at the point of intersection.  The condition $\psi = 0$ on the boundary implies that there is a second term with the eikonal $S_2$ in the sum, which satisfies $S_1=S_2$ and $A_1=-A_2$ on the boundary. As a result tangent derivatives of $S_{1,2}$ on the boundary are also equal and together with Eq.~(\ref{eqeikonal}), this implies that for a non-trivial solution $\hat{n} \cdot \nabla S_2= - \hat{n} \cdot \nabla S_1$. In other words a ray of the eikonal $S_1$ must specularly reflect at the boundary into a ray of another eikonal in the sum, which we label $S_2$.  Hence we know the direction of $\nabla S_2$ at the boundary and can follow it until the next ``reflection'' from the boundary.  Thus each segment of a ray trajectory corresponds to a direction of $\nabla S_m$ for some $m$ in Eq.~(\ref{gradpsi}).  A ray moving linearly in a domain $D$ and specularly reflecting from the boundary describes exactly the same dynamics as a point mass moving on a frictionless ``billiard'' table with boundary walls of shape $\partial D$.  Such dynamical billiards have been studied since Birkhoff in the 1920's as simple dynamical systems which can and typically do exhibit chaotic motion.  Thus the problem of predicting the properties of the vector fields $S_m$ is identical to the problem of the long-time behavior of dynamical billiards.

One property of any such bounded dynamical system (independent of whether it displays chaos) is that any trajectory starting from a point $x_0$ will return to a neighborhood of that point an infinite number of times as $t \rightarrow \infty$ (the Poincar\'e recurrence theorem\cite{poincare1890}).  Therefore we are guaranteed that the ray we followed from $x_0$ in the direction $\hat{p}_1$ will eventually re-enter the neighborhood of size $\lambda$ around $x_0$.  By our previous argument, each linear segment of the ray trajectory, corresponds to one of the directions $\nabla S_m$ and thus when the ray re-enters the neighborhood of $x_0$ for $\nabla \psi$ to be single valued it is necessary that the ray travel in one of the directions $\nabla S_m (\bm{x}_0) = \hat{p}_m$. There can be two categories of ray dynamics: 1)~Although the ray enters the neighborhood of $x_0$ an infinite number of times it only does so in a finite number, $N$ of ray directions.  2)~The number of ray directions grows monotonically with time and tends to infinity as $t \rightarrow \infty$.  We will now show that the general applicability of the eikonal method depends on which category of ray motion occurs.

Let us first consider a billiard $\partial D$ with fully chaotic dynamics.  In the current context ``fully chaotic'' means that for arbitrary choice of $x_0$ and the direction $\hat{p}_1$ (except for sets of measure zero, such as unstable periodic orbits) the distribution of return directions (momenta) is continuous and isotropic as $t \rightarrow \infty$. Therefore the number of terms in an eikonal solution of the form Eq.~(\ref{introuscansatz})  would have to be infinite, contradicting our initial assumption that $N$ was finite.  Thus there do not exist eikonal solutions with finite $N$ for wave equations on domains with fully chaotic ray dynamics. A very closely related point was made by Einstein as early as 1917\cite{einstein17} (he phrased it as the non-existence of a multi-valued vector field defined by the $N$ ``sheets" of the functions $S_m$).  One may ask whether an eikonal solution with an infinite number of terms could be defined; this appears unlikely as the amplitudes for the wavefronts are bounded below by $(\lambda/L)^{1/2}$, where $L$ is the typical linear dimension of $D$, so that only a very special phase relationship between terms would allow such a sum to converge. The essential physics of this breakdown of the eikonal method is that in a chaotic system wave solutions exist but do not have wavefronts which are straight on a scale much larger than a wavelength, hence it is impossible to develop a sensible asymptotic expansion with smooth functions $S_m$.

Returning now to the case of a boundary $\partial D$ for which the distribution of return momenta is always discrete, this means that there exist exactly $N$ ray directions for each point $x_0$ and any choice of $\hat{p}_1$.  In this case the entire spectrum of the wave equation on $\partial D$ can be obtained by an eikonal approximation with $N$ terms of the form Eq.~(\ref{introuscansatz}). The quantized values of $k$ are determined by the conditions that the eikonal only advance in phase by an integer multiple $2 \pi$ upon each return to $x_0$ and hence the solution is single-valued. The correct quantization condition must take into account phase shifts which occurs for rays as they pass caustics. The details of implementing this condition have become know as Einstein-Brillouin-Keller quantization\cite{keller1}.

\begin{figure}[!hbt]
\centering
\subfigure[]
{
\includegraphics[width=0.3\linewidth]{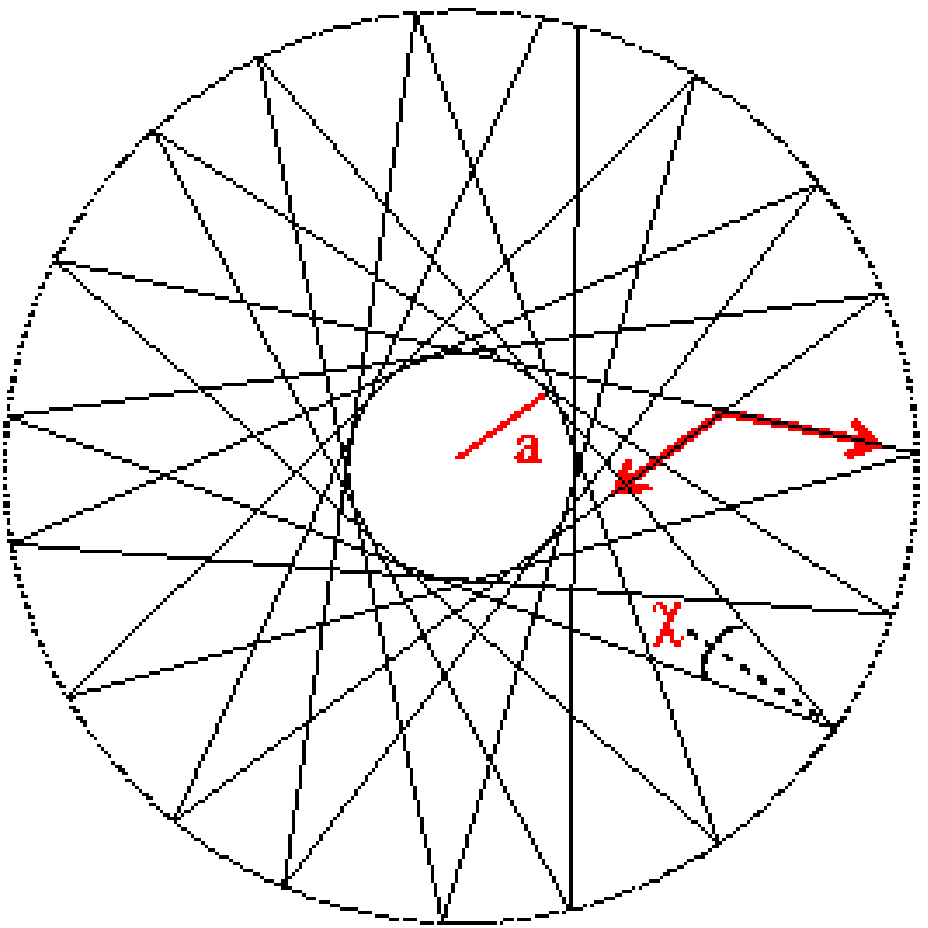}
}
\hspace{3cm}
\subfigure[]
{
\psfrag{l}{ }
\psfrag{p1}{$\hat{p}_1$}
\psfrag{x0}{$x_0$}
\includegraphics[width=0.3\linewidth]{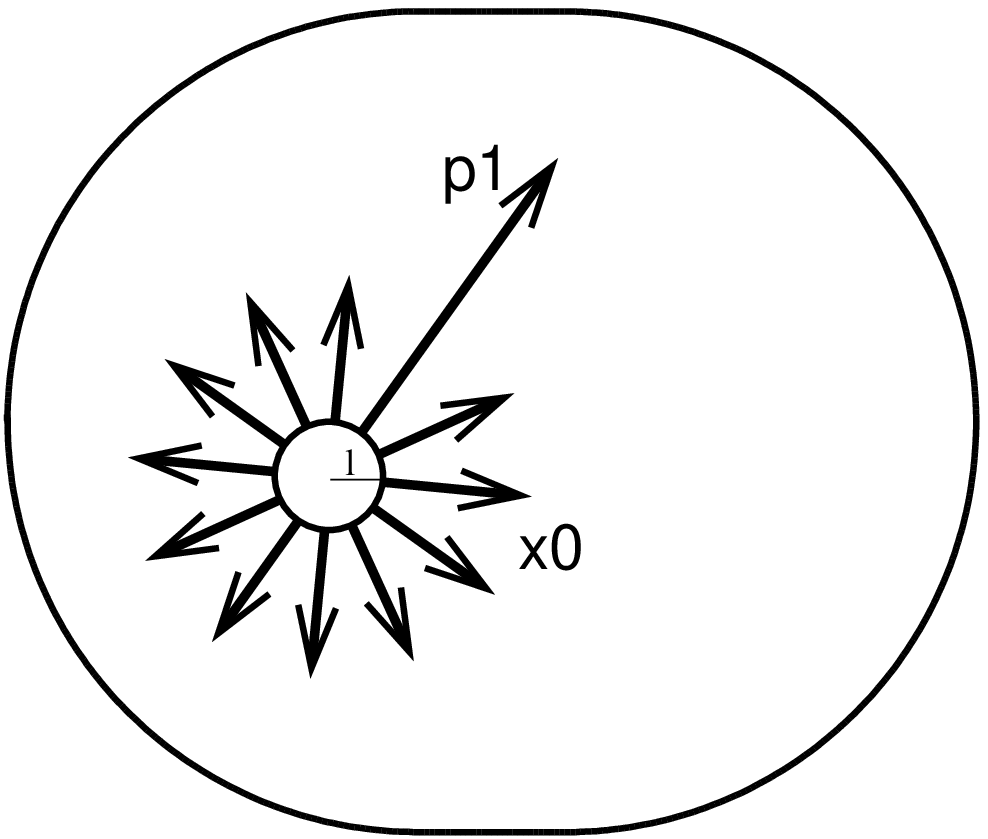}
}
\caption{(a) A typical quasi-periodic ray motion in a circular billiard.
The two possible ray return directions for a specific point $x_0$ and initial direction $\hat{p}_1$ are shown in red. (b) The Bunimovich stadium, consisting of two semi-circles connected by straight segments, for which ray motion is completely chaotic. As the schematic indicates, for any point $x_0$ the ray return directions are infinite, continuously distributed and isotropic, making an eikonal solution impossible.}
\label{figcirclepqp}
\end{figure}

From modern studies of billiard dynamics we know that both of the cases we have just considered are exceptional. The billiards for which eikonal solutions for the entire spectrum is possible are called integrable, and their ray dynamics has one global constant of motion for each degree of freedom. For example in the circular billiard both angular momentum and energy are conserved and for each choice of $x_0$ and direction $\hat{p}_1$ there are exactly two return directions\cite{keller1} (see Fig.~\ref{figcirclepqp}).  While the circle is a good and relevant example here, there are other shapes, such as rectangles and equilateral triangles for which the method also works; obviously these are shapes of very high symmetry. It is also known that an elliptical billiard of any eccentricity is integrable; however this is believed to be the only integrable smooth deformation of a circle\cite{poritsky50,amiran97}. Thus there is a relatively small class of boundaries for which eikonal methods work globally; this point does not seem to be widely appreciated in the optics community.

As already noted, the type of boundary shape which generates continuous return distributions for each choice of $x_0$ and direction $\hat{p}_1$ correspond to completely chaotic billiards and such shapes are also quite rare. No smooth boundary (i.e.\ $\partial D$ for which all derivatives exist) is known to be of this type. A well-known and relevant example for us of such a shape is the stadium billiard, consisting of two semi-circular ``endcaps'' connected by straight sides. Note that the generation of continuous return distributions would fail for a point $x_0$ between the two straight walls if we chose $\hat{p}_1$ perpendicular to the walls generating a (marginally stable) two-bounce periodic orbit passing through $x_0$.  However this choice represents a set of measure zero of the initial conditions in the phase space.  It follows from our above arguments that eikonal methods would fail for the entire spectrum in such a billiard (except a set of measure zero in the short wavelength limit).

The generic dynamics of billiards arises when the boundary is smooth but there is not a second global constant of motion; this is exemplified by the quadrupole billiard we study extensively below (see definition in Eq.~(\ref{quad})). Such a billiard has ``mixed'' dynamics; we shall explain what this means and how it is studied in more detail below.  For such a billiard, depending on the choice of the initial phase space point $(x_0,\hat{p}_1)$, one may get either a finite number $N$ of return directions or an infinite number as $t \rightarrow \infty$.  It is not obvious just from our above arguments that this means that eikonal methods will fail in such a case. We will skip over this point and simply state that in the case of mixed dynamics in principle only a finite fraction of the spectrum could be calculated by eikonal methods. If one can obtain a relatively tractable expression for the locally-conserved quantity which leads to a finite number of return directions $N$, as in the adiabatic approximation of Berry and Robnik\cite{RobnikB85}, then some progress can be made\cite{nature}. However in practice the vector fields $\nabla S_m$ required are usually too complicated to make such an approach tractable.  Thus in practice eikonal methods are not very useful to find solutions of the Helmholtz equation for generic shapes. A related but different analytic method, that of Gaussian optics, can be used to calculate a fraction of the spectrum based on motion near stable periodic orbits.  This method is worked out for dielectric billiards with mixed dynamics in detail in Ref.\cite{Tureci02}.  However both this and the eikonal method fail for a fraction of the spectrum which approaches unity as the chaotic fraction of phase space approaches unity.

Since the traditional analytic methods of optics fail for these systems, what other short wavelength approaches exist? The development of short wavelength approximations for mixed and chaotic systems is precisely the problem of quantum chaos which has been widely studied in atomic, nuclear, solid-state and mathematical physics over the past two decades\cite{gutz_book,haake_book,leshouches89_book, Y2Kproc_book}.  Powerful analytic methods have been developed, but with an essentially different character than eikonal or Gaussian methods (these techniques are typically referred to as {\em semiclassical methods} in the quantum chaos literature).  The analytic methods in quantum chaos theory are all of a statistical character and do not allow one to calculate individual modes.  Instead the methods focus on the fluctuating part of the density of modes and the statistical properties of the spectrum (e.g.\ level-spacing distributions). The results are useful in many contexts, but less useful in the context of optical resonators and micro-lasers for which a single or small set of modes will be selected and one is interested in their emission patterns and Q-values. Therefore it is particularly important to develop efficient numerical methods for calculating the spectrum and modes of such dielectric resonators, and we will discuss our method for doing this in section~(\ref{sect_scatquant1}) below.

We are primarily interested in ARC resonators with mixed billiard dynamics as these shapes lead to resonances with high Q and directional emission. For such resonators, the ray phase space is not fully chaotic, but is highly structured. Moreover the possibility of ray escape decreases the randomizing effect of chaotic motion at $t \rightarrow \infty$.  Therefore using methods which maintain a connection between the wave solutions and the ray phase space is very helpful. We shall describe such a method, known as Husimi projection, in section~(\ref{sect_husimi}) below.

\section{Ray dynamics for generic dielectric resonators}
\label{sect_refbilliard}
Before introducing our numerical method and the Husimi projection method, we review the properties of mixed phase space via the surface of section method in the context of the quadrupole billiard/ARC. This billiard is described by the boundary shape:
\be
R(\phi) = R_0(1 + \eps \cos 2\phi)
\label{quad}
\ee
which in the zero deformation limit $\eps=0$ reduces to a circular billiard, which as we have already noted, is integrable.  Therefore the variation of the parameter $\eps$ starting from zero induces a transition to chaos.  As the perturbation is smooth, various results in dynamical systems (collectively known as Kolmogorov-Arnold-Moser theory) imply that the transition to chaos is gradual\cite{arnold_book,lazutkin_book}. The quadrupole billiard displays the typical behavior characteristic of this transition.  In our initial discussion here we treat the ideal perfectly-reflecting billiard; later we will discuss the role of ray escape in ARCs.

\begin{figure}[!htb]
\psfrag{1}{$1$}
\psfrag{2}{$2$}
\psfrag{3}{$3$}
\psfrag{4}{$4$}
\psfrag{5}{$5$}
\psfrag{-pi}{$-\pi$}
\psfrag{pi}{$\pi$}
\psfrag{nhat}{$\mbf{\hat{\nu}}$}
\psfrag{phi}{$\phi$}
\psfrag{chi}{$\chi$}
\psfrag{sinx}{$\quad\sin\chi_c=\frac{1}{n}$}
\includegraphics[width=\linewidth]{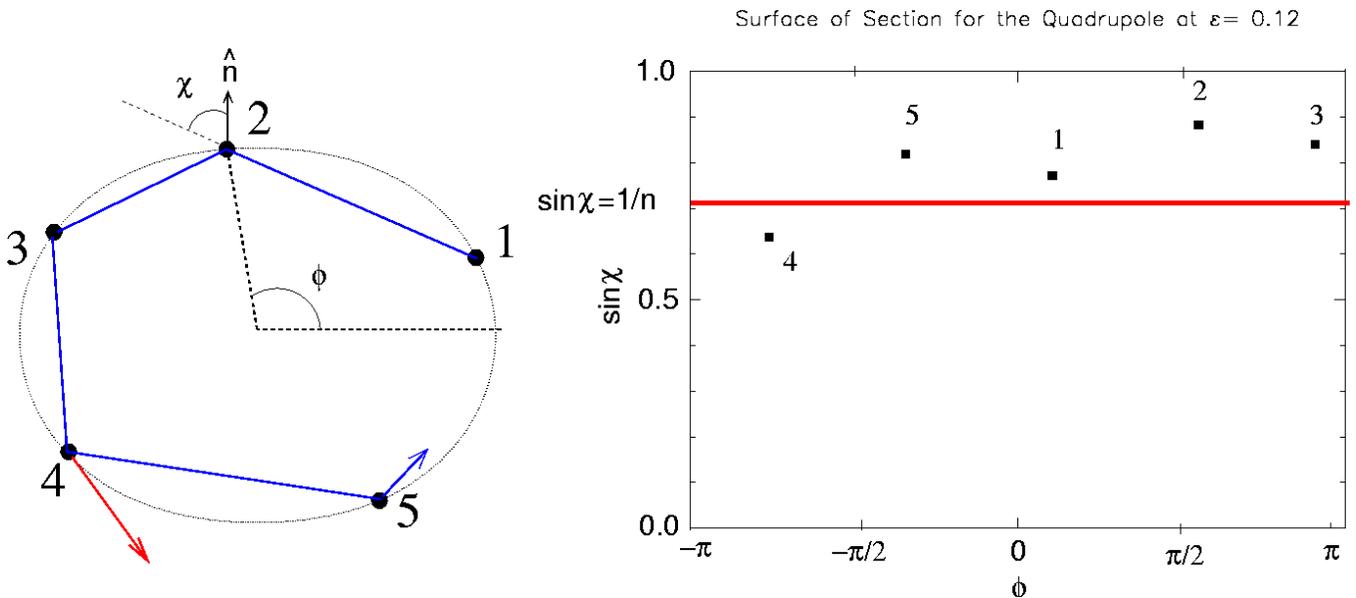}
\caption{The construction of the surface of section plot. Each reflection from the boundary is represented by a point in the SOS recording the angular position of the bounce on the boundary ($\phi$) and the angle of incidence with respect to the local outward pointing normal ($\sin\chi$). For a standary dynamical billiard there is perfect specular reflection and no escape.  For ``dielectric billiards" if $\sin\chi>\sin\chi_c > 1/n$, total internal reflection takes place, but both refraction and reflection according to Fresnel's law results when a bounce point (bounce \#4 in the figure) falls below the ``critical line" (shown in red) $\sin\chi>\sin\chi_c$.   Note that $\sin\chi<0$ correspond to clockwise sense of circulation. We do not plot the $\sin\chi<0$ region as the SOS has reflection symmetry.  Below we will plot the SOS for ideal billiards without escape unless we specify otherwise.}
\label{figdemoSOS}
\end{figure}

When the shape is gradually deformed, it quickly becomes unfeasible to capture the types of ensuing ray motion by standard ray tracing methods in real space. A standard tool of non-linear dynamics, which proves to be very useful in disentangling the dynamical information, is the {\em Poincar{\'e} surface of section} (SOS)\cite{lichtenberg_book,reichl_book}. In this two-dimensional phase-space representation, the internal ray motion is conveniently parametrized by recording the pair of numbers $(\phi_i,\sin\chi_i)$ at each  reflection $i$, where $\phi_i$ is the polar angle denoting the position of the $i$th reflection on the boundary and $\sin\chi_i$ is the corresponding angle of incidence of the ray at that position (see Fig.~\ref{figdemoSOS}). Each initial point is then evolved in time through the iteration of the SOS map $i\ra i+1$, resulting in  basically two general classes of distributions. If the iteration results in a one-dimensional distribution (an {\em invariant curve}), the motion represented is {\em regular}. On the other hand exploration of a two-dimensional region is the signature of {\em chaotic} motion.

\begin{figure}[!hbt]
\centering
\psfrag{epsilon0}{\footnotesize $\varepsilon=0.0$}
\psfrag{epsilon05}{\footnotesize $\varepsilon=0.05$}
\psfrag{epsilon11}{\footnotesize $\varepsilon=0.11$}
\psfrag{epsilon18}{\footnotesize $\varepsilon=0.18$}
\includegraphics[width=\linewidth]{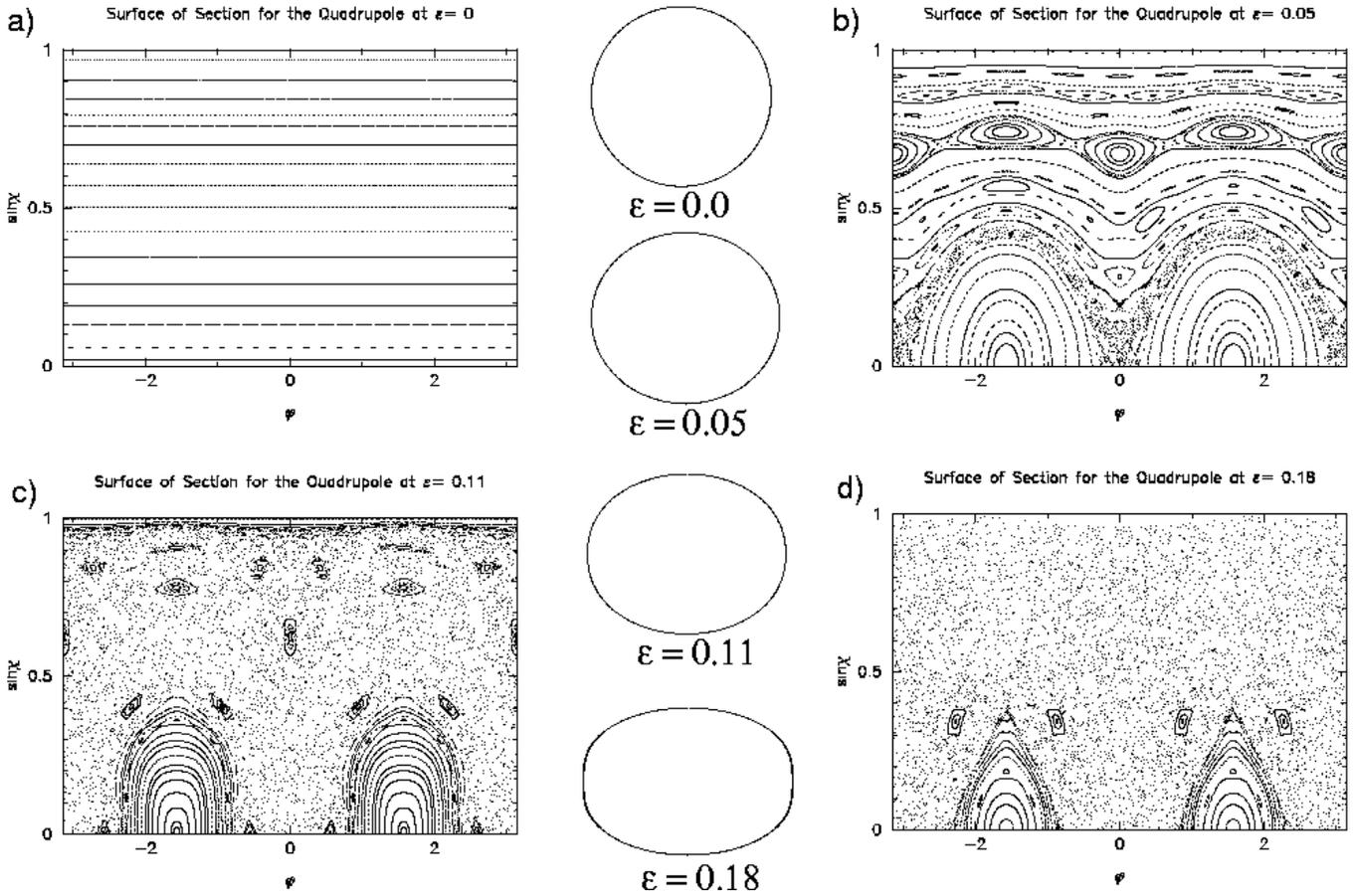}
\caption{The SOS of a quadrupole at fractional deformations $\eps= 0,0.05,0.11,0.18$. The closed curves and the curves crossing the SOS represent two types of regular motion, motion near a stable periodic orbit and quasi-periodic motion respectively. The regions of scattered points represent chaotic portions of phase space. A single trajectory in this ``chaotic component" will explore the entire chaotic region. With increasing deformation the chaotic component of the SOS (scattered points) grows with respect to regular components and is already dominant at 11 \% deformation.  Note in (b) the separatrix region associated with the two-bounce unstable orbit along the major axis where the transition to chaotic motion sets in first.}
\label{figSOScol}
\end{figure}

The transition to ray chaos in the quadrupole billiard is illustrated in Fig.~\ref{figSOScol}. At zero deformation the conservation of $\sin\chi$ results in straight line trajectories throughout the SOS and we have globally regular motion. These are the well-known whispering gallery (WG) orbits for $\sin \chi > 1/n$. As the deformation is increased (see Fig.~\ref{figSOScol}) chaotic motion appears (the areas of scattered points in Fig.~\ref{figSOScol}) and a given initial condition explores a larger range of values of $\sin \chi$. Simultaneously, islands of stable motion emerge (closed curves in Fig.~\ref{figSOScol}), but there also exist extended ``KAM curves"\cite{lazutkin_book} (open curves in Fig.~\ref{figSOScol}), which describe a deformed WG-like motion close to the perimeter of the boundary. These islands and KAM curves cannot be crossed by chaotic trajectories in the SOS. As the transition to chaos ensues, a crucial role is played by the {\em periodic orbits} (POs), which appear as fixed points of the SOS map. The local structure of the islands and chaotic layers can be understood through the periodic orbits which they contain. Thus, the center of each island contains a stable fixed point, and close to each stable fixed point the invariant curves form a family of rotated ellipses. The Birkhoff fixed point theorem\cite{lichtenberg_book} guarantees that each stable fixed point has an unstable partner, which resides on the intersection of separatrix curves surrounding the elliptic manifolds. Chaotic motion sets in at separatrix regions first, and with increasing deformation pervades larger and larger regions of the SOS. Already at $\eps=0.1$, much of the phase space is chaotic and a typical initial condition in the chaotic sea explores a large range of $\sin\chi$, eventually reversing its sense of rotation.

\section{Formulation of the resonance problem}
\label{sect_resprob}
A dielectric resonator is significantly different from the closed (Dirichlet) problem due to its openness. In contrast to ideal metallic cavities which possess normal modes at discrete real frequencies, dielectric resonators are characterized by a discrete set of {\em quasi-bound modes}\cite{LeungLY94,ChingLvSTY98}, or {\em resonances}. As a result, the quasi-bound modes of a resonator are characterized by a frequency $\omega=ck$ and a lifetime $\tau$, where $c$ is the speed of light and $k=2\pi/\lambda$ is the wavevector in vacuum. Experiments on resonators fall into two broad categories, and the presence of quasi-bound modes are manifested differently in these two situations.

In scattering experiments, an incoming field produced by a source in the {\em farfield} (spatial infinity) gives rise to an outgoing field which represents the response of the resonator, as measured by an ideal detector in the farfield. In the ideal case, where absorption is absent, this corresponds to a situation where energy is conserved and hence in this situation the EM field has a real frequency, $\omega$, which is arbitrary and set by the source. In emission experiments, on the other hand, there is no incoming field, but only an outgoing field. As a result, energy is depleted from the system, and this process is characterized by decay. The simplest mathematical description of these two experiments correspond to the solution of the wave-equation (which is derived from the Maxwell's equations as described in section~(\ref{sect_wavereduct}))
\be
\left(\bm{\nabla}^2 - \frac{n^2(\bm{x})}{c^2}\pder{2}{}{t} \right)
\Psi (\bm{x},t) = 0
\ee
where the solutions have the separable, time-harmonic dependence
\be
\Psi (\bm{x},t) = \psi(\bm{x}) \ex{i\omega t}
\label{tdepndtpsi}
\ee
so that $\psi(\bm{x})$ obeys the {\em Helmholtz equation}
\be
\left(\bm{\nabla}^2 + n^2(\bm{x})k^2 \right)\psi(\bm{x}) = 0
\label{hheqintro2}
\ee
Here, $n(\bm{x})$ represents the index of refraction. In general, one can define a complete set of incoming $\{\psi^{(-)}_{\mu}(k;\bm{x})\}$ and outgoing modes $\{\psi^{(+)}_{\mu}(k;\bm{x})\}$ at a given $k$, in the absence of the resonator. The exact form of these sets is dictated by convenience, and in the present discussion we will employ the cylindrical harmonics.

\begin{figure}[!hbt]
\psfrag{iphi}{$I(170^{\circ})$}
\psfrag{rekR}{$\re{kR}$}
\psfrag{imkR}{$\im{kR}$}
\centering
\includegraphics[width=0.7\linewidth]{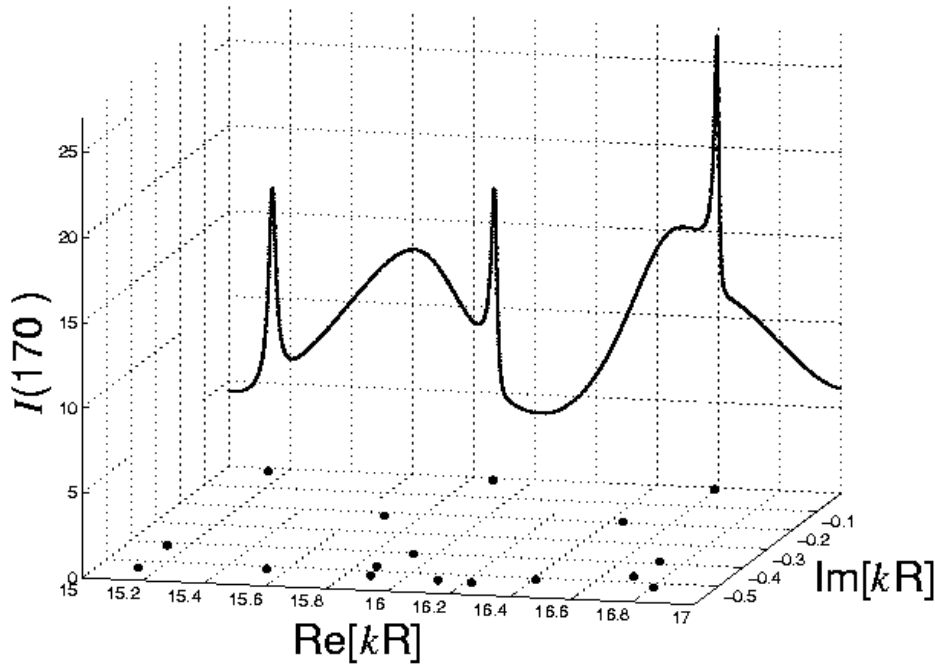}
\caption{A comparison of scattering and emission pictures for quasi-bound modes. Variation of the intensity scattered off a dielectric circular cylinder with the wavenumber $k$ of an incoming plane-wave is plotted on the back panel ($I-\re{kR}$ plane). The intensity is observed at $170^\circ$ with respect to the incoming wave direction; it would look significantly different in another direction due to interference with the incident beam. The complex quasi-bound mode frequencies are plotted on the $\re{kR}-\im{kR}$ plane. Notice that the most prominent peaks in scattering intensity are found at the values of $k$ where a quasi-bound mode frequency is closest to the real-axis. These are the long-lived resonances of the cavity. Also visible is the contribution of resonances with shorter lifetimes (higher values of $\im{kR}$) to broader peaks and the scattering background.}
\label{figscattering}
\end{figure}

The two experimental situations at this point are distinguished by two different boundary conditions in the farfield. The scattering experiment corresponds to the boundary condition
\be
\psi(\bm{x}) \sim \psi^{(-)}_{\mu}(k;\bm{x}) +
\sum_{\nu}S_{\mu\nu}(k) \psi^{(+)}_{\nu}(k;\bm{x}),  \qquad
|\bm{x}|\ra \infty
\label{introscatans}
\ee
and experimentally, it is the scattering matrix $S_{\mu\nu}(k)$, which contains the information measured by the farfield detector. In the typical case $S_{\mu\nu}(k)$ will display sharp peaks at a discrete set of real wavevectors $k_i$ (in case of isolated resonances; see the back panel ($\re{kR}$-$I$ plane) on Fig.~\ref{figscattering}). This is the signature of {\em long-lived} quasi-bound modes with frequency $\omega=ck_i$; their lifetimes $\tau$ are encoded in the functional form of the peaks, which in general is of Fano shape, with direction-dependent parameters.  This makes scattering boundary conditions less convenient for the extraction of the quasi-bound mode structure.

The emission experiments are modeled by the outgoing wave boundary conditions at infinity
\be
\psi^{(i)}(\bm{x}) \sim \sum_{\nu}\gamma_{\nu}(k_i)
\psi^{(+)}_{\nu}(k_i;\bm{x}), \qquad |\bm{x}|\ra \infty
\label{introqbm}
\ee
This form at infinity does not permit solution for any real $k$ as it manifestly violates current conservation. Instead the solutions of Eq.~(\ref{hheqintro2}), indexed by $i$, exist only at discrete complex wavevectors $k_i=\kappa_i + i\Gamma_i$. The connection to quasi-bound modes is then direct; the real part gives the quasi-bound mode frequency $\omega_i=c\kappa_i$ and the imaginary part represents the lifetime of the mode, $\tau_i = 1/c\Gamma_i$. Here, we will use the radiation boundary conditions exclusively, and quote the dimensionless complex variable $kR$ instead, where $R$ is the mean radius of the resonator. In an active medium, one may think of these resonances as being pulled up to real wave-vectors by the gain. Although the description of actual (stationary) laser modes requires the solution of a non-linear wave-equation\cite{tureci_thesis}, we will focus here on the problem of linear resonances, an approximation which is often used in laser theory\cite{siegman_book}.

The relation between the linear emission and scattering picture is easily visualized in the extended complex wavevector space of the scattering matrix $S_{\mu\nu}(k)$, depicted in Fig.~\ref{figscattering}. The discrete quasi-bound wavevectors $k_i$ are the {\em poles} of $S_{\mu\nu}(k)$. As can be seen from the figure, in general there are multiple quasi-bound modes contributing to a given resonance peak, but the quasi-bound modes which are closest to the real-axis lead to the sharpest peaks (some of which might not even be resolved in the scattering profile). Note that via Eq.~(\ref{tdepndtpsi}), the quasi-bound mode solutions damp in time. An important experimental value often quoted is the Q-value of a resonator, which is defined by the number of cycles of the optical field at frequency $\omega$ to decay to half of its value, and thus can be related to quasi-bound mode parameters by the relation $Q=\omega\tau= -2\re{kR}/|\im{kR}|$.

It is possible to generalize eikonal theory to calculate the quasi-bound states of complex $k$ of dielectric cavities of integrable shape\cite{tureci_thesis}. The dielectric boundary conditions then reduce to ray trajectories which still propagate on straight lines and undergo specular reflection, much like in the case of billiards. The additional feature is that dielectric billiards exhibit {\em ray splitting} at the boundary, and give rise to both a refracted and reflected ray with amplitude and direction obtained from the application of the laws of Snell and Fresnel for a flat dielectric boundary. The transport equations for the amplitude have to be supplemented by an additional complex multiplicative factor at each encounter with the boundary. The practical implication is that the ray motion as displayed on the SOS can be used for the dielectric problem, when augmented with an escape condition. The escape probability will be exponentially small in wavenumber $k$ for angles of incidence above the critical angle $\chi_c = \sin^{-1} 1/n$, since its due to a tunneling-like process\cite{noeckel_thesis,tureci_thesis}. This condition is demarcated by the line $\sin\chi=\sin\chi_c$ in the SOS; any ray falling below this line refracts out with the probability given by the local Fresnel law of refraction (assuming a TM mode):
\be
T(\sin\chi) = \frac{2\sqrt{1-\sin^2\chi}}{\sqrt{1-\sin^2\chi} +
\sqrt{\sin^2\chi_c - \sin^2\chi}}
\label{introtrans}
\ee
providing a classical loss mechanism and leading to finite lifetimes of the corresponding modes.  Thus the openness of dielectric resonators is not the cause of the failure of standard methods; these methods fail for chaotic resonator shapes for the same reasons of dynamical complexity discussed in section~\ref{sect_torus_quant} for the Dirichlet case. We now discuss the reduction of the wave equation for dielectric resonators to the Helmholtz equation to lay the groundwork for the numerical method of section~\ref{sect_scatquant1}, which allows us to solve for the resonances of chaotic shapes.

\section{Reduction of Maxwell's equations}
\label{sect_wavereduct}

\begin{figure}[!hbt]
\psfrag{DD}{$\partial D$}
\psfrag{E}{$\bm{E}$}
\centering
\includegraphics[width=0.6\linewidth]{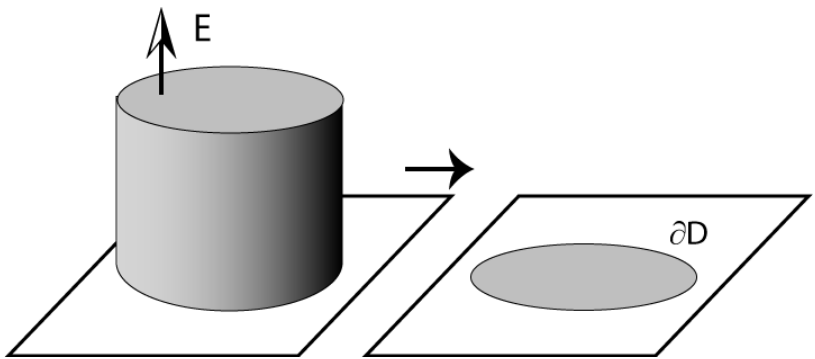}
\caption{Illustration of the reduction of the Maxwell equation for an infinite dielectric rod of general cross-section to the 2D Helmholtz equation for the TM case (E field parallel to axis) and $k_{\parallel}=k_z =0$.}
\label{figcylinder}
\end{figure}

Consider the problem of excitation of electromagnetic waves in an infinite dielectric rod of arbitrary cross-section (see schematics in Fig.~\ref{figcylinder}), which is extended along the z-axis. In practical situations, the structure is of finite extent and there are planar end-caps which makes it truly a resonator. In other cases, it's a fiber-optic cable of practically infinite extent. In any case, we will for now assume translational symmetry along z-axis, and we will show later that this is a perfectly valid assumption for the modes of relevance to us.

Assuming harmonic time-dependence of the fields and no surface currents and charges, Maxwell's equations for the system yield the Helmholtz equations:
\be
\left(\nabla^2 + n^2(\bm{x}) k^2\right)
\left\{
\ba{c}
\bm{E}\\
\bm{B}
\ea
\right\}
=0
\label{vectorhheqn}
\ee
where $k=2\pi/\lambda=\omega/c$ is the wavevector in vacuum; $n=\sqrt{\mu\eps}$ is the index of refraction, $\mu$ is the permeability and $\eps$ is the dielectric constant of the medium, which are in general function of position. We will assume $\mu=1$, so that $n^2=\eps$.

The translational symmetry along the $z$-axis allows us to express the $z$-variation of the fields as
\be
\bm{E}(\bm{x}) = \bm{E}(x,y)\ex{ink_z z}
\ee
Following Jackson\cite{jackson_book}, we separate the fields and operators into components parallel and transverse to the $z$-axis and write out the transverse projection of curl equations:
\bea
ink_z\bm{E}_{\perp} + ik\bm{z \times \bm{B}_{\perp}} & = &
\bm{\nabla_{\perp}}E_z \\
ink_z\bm{B}_{\perp} - ink\bm{z \times \bm{E}_{\perp}} & = &
\bm{\nabla_{\perp}}B_z
\label{max2eqs}
\eea
It's evident from these four (scalar) equations that $E_z$ and $B_z$ are the fundamental fields we should be after, and that once they are determined we can solve for $\bm{E}_{\perp}$ and $\bm{B}_{\perp}$. Thus, the Maxwell's equations themselves completely decouple, which was already obvious from Eq.~(\ref{vectorhheqn}). The actual complication of solving the vector Helmholtz equation stems from the fact that the {\em boundary conditions} are coupled. The Maxwell boundary conditions are
\bea
\hat{\bm{\nu}}\times(\bm{E}_1 -\bm{E}_2) = 0, \quad \hat{\bm{\nu}}
\cdot (n_1^2\bm{E}_1 -n_2^2\bm{E}_2) = 0 \\
\hat{\bm{\nu}} \times (\bm{B}_1 -\bm{B}_2) = 0, \qquad
\hat{\bm{\nu}} \cdot (\bm{B}_1 -\bm{B}_2) = 0
\eea
in the absence of surface currents and charges and for a linear, isotropic medium. The subscripts denote the media on respective sides of the interface. $\hat{\bm{\nu}}$ is the unit normal on the interface, pointing towards out from the cylinder. We will assume $n_1=n>n_2=1$. Note that these are six conditions altogether. Focusing on the scalar fields $E_z$, $B_z$, we have from the equations involving the cross-product with $\hat{\nu}$, $E_{1z} = E_{2z}$ and $B_{1z}=B_{2z}$. Another pair of boundary conditions can be found by projecting Eq.~(\ref{max2eqs}) onto the diad $(\hat{\bm{\nu}},\hat{\bm{s}})$ defined on the boundary of the cross-section $\partial D$
\bea
\pder{}{E_{1z}}{s} - \pder{}{E_{2z}}{s} &=& -\frac{k_z}{k} \left(
\frac{1}{n}\pder{}{B_{1z}}{\nu} - \pder{}{B_{2z}}{\nu} \right)\\
\pder{}{E_{1z}}{\nu} - \pder{}{E_{2z}}{\nu} &=& \frac{k_z}{k} \left(
\frac{1}{n}\pder{}{B_{1z}}{s} - \pder{}{B_{2z}}{s} \right)
\label{maxgenbcs}
\eea
We are interested in the long-lived modes of the resonator. Modes with a finite $k_z$ correspond in short wavelength limit to rays which spiral up and down along the cylinder walls and escape through the end-caps by refracting out. Thus under most circumstances the longest lived modes have $k_z \approx 0$, and correspond to modes which are effectively two-dimensional, i.e.\ can be expressed by dynamics of rays on the cross-sectional plane. In that case, the boundary conditions Eq.~(\ref{maxgenbcs}) also become diagonal and we have a complete decoupling. We will choose to work with $\psi_i(x,y) = E_{iz}(x,y)$, corresponding to TM polarized fields, for which the problem reduces to the two-dimensional Helmholtz equation for the scalar field $\psi$ with continuity conditions
\bea
\left( \nabla^2_{\perp} + n_i^2 k^2 \right) \psi_i(x,y) = 0 &
\label{hheqits}\\ \psi_1 |_{\partial D} = \psi_2 |_{\partial D},
\quad \pder{}{\psi_1}{\nu}|_{\partial D} =
\pder{}{\psi_2}{\nu}|_{\partial D} &
\label{hheqpbcs}
\eea
Note that this boundary value problem is equivalent to that of the stationary Schr\"odinger equation of quantum mechanics. Hereafter, we will drop all references to the original three-dimensional and vector character of the problem and work with Eqs.~(\ref{hheqits})-(\ref{hheqpbcs}).

\section{Scattering Quantization-Philosophy and Methodology}
\label{sect_scatquant1}
In this section we will describe a numerical method to solve Eq.~(\ref{hheqpbcs}), which is both efficient and physically appealing. Our approach is a generalization to open systems (specifically, dielectric resonators) of the {\em scattering quantization approach} to quantum billiards\cite{doron92,DietzEPSU95}. This approach is based on the observation that every quantum billiard interior problem (Helmholtz equation for a bounded region with Dirichlet/Neumann boundary conditions) can be viewed as a scattering problem, and the spectrum can be uniquely deduced from the knowledge of the corresponding scattering operator. In the case of closed systems, the internal scattering problem can be mapped rigorously to an external scattering problem\cite{EckmannP95}, and the resulting (exact) scattering matrix is unitary. For the dielectric resonator problem with radiation boundary conditions, we will see that the corresponding scattering operator is inherently non-unitary, reflecting the physical fact that we are dealing with a leaky system. Thus we will define below a new ``S-matrix" which is non-unitary and distinct from the true S-matrix describing external scattering from the system. We retain the terminology ``S-matrix" nonetheless because of the conceptual similarity to the quantum billiard method of \cite{doron92,DietzEPSU95}. The generalization of this approach to dielectric billiards was first made in Ref.\cite{evgeni2}, however without the efficient algorithm presented below.

We assume that the resonator is bounded by the interface $\partial D$ of the form $r = R(\phi)$, where $R(\phi)$ is some smooth deformation of the boundary such that there exists only one point of the boundary for each angle $\phi$. We decompose the internal and external fields into cylindrical harmonics with a constant $k$
\[
\parbox{0.7\linewidth}{
\begin{eqnarray*}
\psi_1(r,\phi) &=& \sum_{m=-\infty}^{\infty} \left( \alpha_m
\beshp{m}{nkr} + \beta_m \beshm{m}{nkr} \right) \ex{im\phi} \\
\psi_2(r,\phi) &=& \sum_{m=-\infty}^{\infty} \left( \gamma_m
\beshp{m}{kr} + \delta_m \beshm{m}{kr} \right) \ex{im\phi}
\end{eqnarray*}
} \hfill
\parbox{0.3\linewidth}{
\setlength{\jot}{23pt}
\bea
r&<&R(\phi) \label{intwf} \\
r&>&R(\phi) \label{extwf}
\eea
}
\]
Each of the terms
\be
\psi_m^{\pm}(r,\phi) = \beshpm{m}{nkr}\ex{im\phi}
\label{eqangmomrep}
\ee
in the sum is a solution of the appropriate (interior or exterior) Helmholtz equation, but does not satisfy the matching conditions by itself. Note that $\{\psi_m^{\pm}\}$ forms a {\em normal basis} in the infinite space. Owing to the completeness of this basis, the expansion is exact for $r<R_{min}$ and $r>R_{max}$ as long as the sum runs over an {\em infinite} number of terms, where $R_{min}$ and $R_{max}$ are the lower and upper bounds of $R(\phi)$ respectively. The assumption that the expansions can be analytically continued to the region $R_{min}<r<R_{max}$ is known as the Rayleigh hypothesis\cite{rayleigh07}. It has been shown\cite{berg79} that for a family of deformations parametrized by $\eps$, there is typically a critical deformation $\eps_c$, beyond which the hypothesis breaks down because the expansion ceases to be analytic in the region $R_{min}<r<R_{max}$. For the deformations Eq.~(\ref{quad}), this happens long after the shape becomes concave; we are not interested in this regime. Although this issue seems thus to be resolved, we shall see that precursors of the non-convergence emerge in the form of numerical instabilities for $\eps < \eps_c$.

We will assume that $\delta_m=0$ (no incoming waves), thus confining our attention to quasi-bound modes. Turning to the interior expansion, the regularity of the solution at the origin requires that we take $\alpha_m = \beta_m$, but we will not implement this condition at this stage. The continuity conditions Eq.~(\ref{hheqpbcs}) give us further relations among the remaining coefficients:
\bea
\psi_1(\phi,R(\phi)) &=& \psi_2(\phi,R(\phi)) \label{fieldseq} \\
\pder{}{\psi_1}{r}\left|_{\phi, R(\phi)} \right. &=&
\pder{}{\psi_2}{r}\left|_{\phi,R(\phi)}\right.
\label{derseq}
\eea
In Eq.~(\ref{derseq}), we have replaced the normal derivative condition by the radial derivative condition, because Eq.~(\ref{fieldseq}) shows that the tangential derivatives are also continuous. Note that this latter set of equations containing radial derivatives is equivalent to the set of equations~(\ref{hheqpbcs}) using normal derivatives.

These conditions can be written out as
\bea
\sum_{m=-\infty}^{\infty} \left( \alpha_m \beshp{m}{nkR(\phi)} +
\beta_m \beshm{m}{nkR(\phi)} \right) \ex{im\phi} &=&
\sum_{m=-\infty}^{\infty} \gamma_m \beshp{m}{kR(\phi)} \ex{im\phi}
\label{Hprecond1} \\
n\sum_{m=-\infty}^{\infty} \left( \alpha_m \mbox{H}_m^{+
\prime}(nkR(\phi)) + \beta_m  \mbox{H}_m^{- \prime}(nkR(\phi))\right)
\ex{im\phi} &=& \sum_{m=-\infty}^{\infty}  \gamma_m \mbox{H}_m^{+
\prime}(kR(\phi)) \ex{im\phi}\label{Hprecond2}
\eea
We multiply both sides by $w_n(\phi)\ex{-in\phi}$ and integrate with respect to $\phi$ to get a matrix equation for the coefficient vectors $\bra{\alpha}$, $\bra{\beta}$ and $\bra{\gamma}$
\bea
{\cal H}_1^+ \bra{\alpha} + {\cal H}_1^- \bra{\beta} &=& {\cal H}_2^+ \bra{\gamma} \label{Hcond1}\\
{\cal DH}_1^+ \bra{\alpha} + {\cal DH}_1^- \bra{\beta} &=& \frac{1}{n} {\cal DH}_2^+
\bra{\gamma} \label{Hcond2}
\eea
Various choices of the weight function $w(\phi)$ are possible\cite{evgeni2}; here we choose $w(\phi)=1$. The matrices in Eq.~(\ref{Hcond1},~\ref{Hcond2}) are defined by
\bea
\left[ {\cal H}_j^{\pm} \right]_{lm} &=& \int_{0}^{2\pi} d\phi
\,\beshpm{m}{n_j k R(\phi)} \ex{i(m-l)\phi} \\
\left[ {\cal DH}_j^{\pm} \right]_{lm} &=& \int_{0}^{2\pi} d\phi
\,\mbox{H}_m^{\pm \prime}(n_j k R(\phi)) \ex{i(m-l)\phi}
\eea
Eliminating $\bra{\gamma}$ between Eq.~(\ref{Hcond1}) and Eq.~(\ref{Hcond2}), we obtain
\be
\mcal{S}(k) \bra{\alpha} = \bra{\beta}
\ee
where the matrix $\mcal{S}(k)$ is given by
\be
\mcal{S}(k) =
\left[n({\cal DH}_2^+)^{-1}{\cal DH}_1^- - ({\cal H}_2^+)^{-1}{\cal H}_1^- \right]^{-1}
\left[({\cal H}_2^+)^{-1}{\cal H}_1^+ - n({\cal DH}_2^+)^{-1}{\cal DH}_1^+ \right]
\label{openSmatform}
\ee
As noted earlier, this S-matrix is different from the standard external scattering matrix introduced in Eq.~(\ref{introscatans}). It is straightforward to check that, for real $k$, $\mcal{S}(k)$ is non-unitary. Consider now the eigenvalue problem of $\mcal{S}(k)$
\be
\mcal{S}(k) \bra{\alpha} = \ex{i\varphi} \bra{\alpha},
\label{evalproblem}
\ee
where for real $k$ the phase $\varphi$ is complex. Once we find a complex $k_q$ where one (or several) of the $\varphi$ is a multiple of $2\pi$, we have $\bra{\alpha} = \bra{\beta}$, which is exactly the condition of regularity at the origin. This is the quantization condition which will provide us with the quantized eigenvalues and eigenvectors $(k_q,\bra{\alpha^{(q)}})$ which allow us to construct the resonant solutions of the interior and exterior problem we set out to find. This condition is often expressed in terms of the secular function $\zeta(k)$\cite{doron92} given by
\be
\zeta(k) = \det [1-\mcal{S}(k)]
\label{eqsecfn}
\ee
The spectrum is obtained as the zeros of the secular equation $\zeta(k) = 0$. As noted, the values $k_q$ for which we obtain a unit eigenvalue and the secular function Eq.~(\ref{eqsecfn}) has a root, is always complex and the eigenvalues of $\mcal{S}(k)$ are not pure phases, $\varphi \in \mathbb{C}$. The practical upshot of this is that this requires a two-dimensional root-search for the equation $\zeta (k) = 0$. An often employed numerical procedure involves a sweep in the complex $k$-plane of the singular values of the operator $T(k)=1-\mcal{S}(k)$, with proper care of the numerical null-space of $T(k)$\cite{alex_thesis}. This requires several calculations of the entries of $T(k)$ and its singular value decomposition per quantized state. In the next section, we will represent an efficient root-finding method, which ideally requires {\em two} diagonalizations per $nkR$ quantized states. Before doing that however, it's worthwhile to investigate the structure of $\mcal{S}(k)$ based on simple physical considerations.
\begin{figure}[t]
\psfrag{alpha}{$\bra{\beta}$}
\psfrag{gamma}{$C$}
\psfrag{d}{$\delta$}
\psfrag{Sdalpha}{$S_{\delta}\bra{\alpha}$}
\psfrag{SSalpha}{$S_{\delta}\mcal{S}\bra{\alpha}$}
\centering
\includegraphics[width=0.5\linewidth]{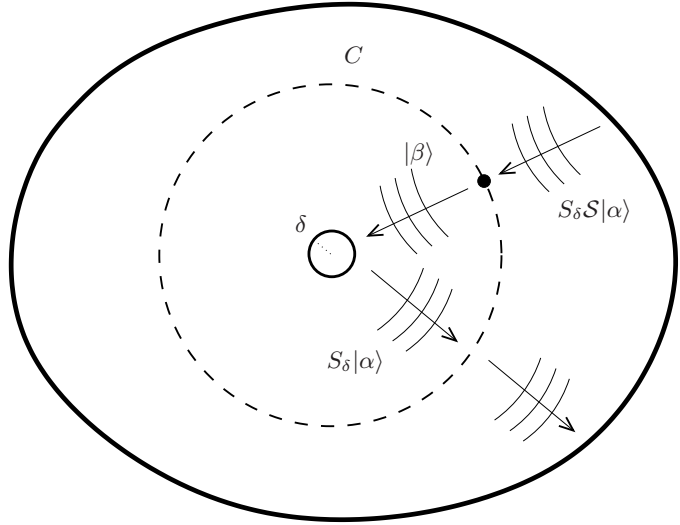}
\caption{Schematics describing the quantum Poincar{\'e} mapping
induced by the internal scattering operator, see discussion in text.}
\label{figsqschem}
\end{figure}

A physical interpretation of the internal scattering operator $\mcal{S}(k)$ and its eigenvectors can be given even off-quantization ($\varphi(k) \neq 2\pi$)\cite{klakow96,frischat97}. We can visualize this approach in our case by dividing the interior of the resonator into two subdomains joined along the curve $C$, which we take to be circle of radius $R_{C}\stackrel{<}{\sim}R_{min}$, and considering it as a boundary at the junction of two back-to-back scattering systems. We furthermore introduce a tiny metallic inclusion of radius $\delta$ at the origin (this is introduced for the sake of the argument and can be omitted). This is our first scattering system, which scatters an incoming wave $\bra{\beta}$ into $\bra{\alpha}$ via the scattering operator $S_{\delta}$
\be
\bra{\beta} =  S_{\delta}(k) \bra{\alpha}
\ee
$S_{\delta} (k)$ is exactly the exterior scattering operator for a metallic circle (immersed in a medium with index of refraction $n$):
\be
[S_{\delta}(k)]_{mm^{\prime}} =
-\frac{\beshm{m}{nk\delta}}{\beshp{m}{nk\delta}} \delta_{mm^{\prime}}
\ee
The second scattering system is the boundary itself, scattering an incoming wave (with respect to the boundary) $\bra{\alpha}$ into $\bra{\beta}$, and the scattering operator for this system is simply $\mcal{S}(k)$ whose form is given in Eq.~(\ref{openSmatform}). Consider now a whole cycle, starting with the state $\bra{\alpha}$ on $C$, being first scattered off the tiny circle, then from the boundary returning to $C$ again (see Fig.~\ref{figsqschem}). The resulting scattered vector is $\mcal{S}\cdot S_{\delta}\bra{\alpha}$. Now, as $k\delta \ra 0$, we have $S_\delta \ra 1$, and the resulting scattered vector is $\mcal{S}\bra{\alpha}$. Because the individual normal modes $\psi_m^{\pm}$ in our expansion correspond to ray trajectories which have a well-defined angular momentum $\sin\chi=\frac{m}{nkR_C}$, the mapping $\mcal{S}\bra{\alpha}$ can be interpreted as a wave analogue of the Poincar\'e SOS mapping on the section $C$, parametrized by $(\phi,\sin\chi)$. This link has been fruitfully used to obtain short wavelength forms of the scattering  operator $\mcal{S}(k)$, for various closed systems\cite{klakow96}. We will not pursue this approach here, but will make use of this visualization to develop a meaningful truncation scheme for a numerical implementation of our method.

\begin{figure}[!t]
\centering
\includegraphics[width=0.5\linewidth]{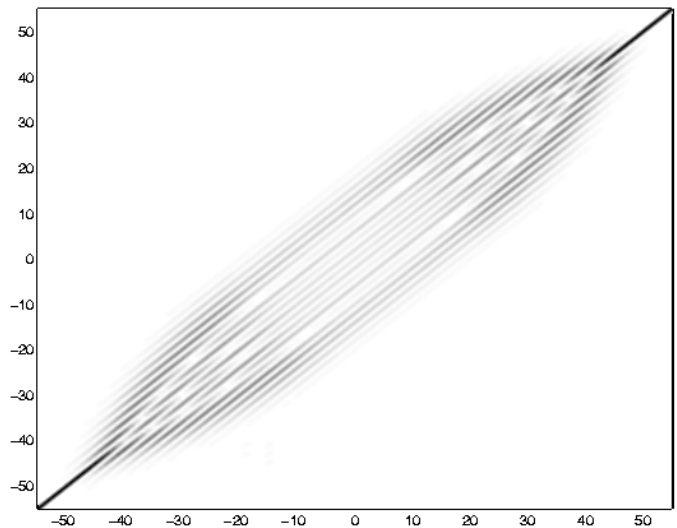}
\caption{A gray-scale representation of the scattering matrix Eq.~(\ref{openSmatform}), calculated for a quadrupolar resonator at $\eps=0.1$ deformation, $n=2.5$ and $nkR=40$. The number of evanescent channels used in the calculation is $\Lambda_{ev}=15$. Note the strong diagonal form for $|m|>nkR$. The spread around the diagonal is proportional to the deformation. Here the internal scattering couples approximately $20$ angular momentum modes.}
\label{figSmatrix}
\end{figure}

First of all, at a given $k$, an angular momentum eigenstate $\psi_m^{\pm}$, for which $m>nkR_{max}$ is a closed channel for  the internal scattering system, because it corresponds to classical motion with a circular caustic of radius larger than $R_{max}$. Such channels are called {\em evanescent}, and are not not expected to be scattered significantly. In fact, a plot of the matrix $\mcal{S}(k)$ in Fig.~\ref{figSmatrix} reveals that as $m$ grows beyond a critical value $m_c \approx nkR_{max}$, the scattering matrix becomes strongly diagonal i.e. $[\mcal{S}(k)]_{mm^{\prime}} \approx \delta_{mm^{\prime}}$ for $|m|,|m^{\prime}| > m_c$. Furthermore, there is a transition region $nkR_{min}<|m|,|m^{\prime}|<nkR_{max}$, where the matrix is heading towards diagonality, and this region corresponds to evanescent components which undergo an enhanced scattering because they overlap significantly with only certain regions of the resonator. This region grows with the deformation of the resonator, and consequently, $\Lambda_{ev}$ evanecent channels have to be included in the number $\Lambda$ of (positive) channels contributing to a given internal scattering matrix. Deonoting the critical matrix size at the evanescent channel boundary by $\Lambda_{sc} = [\![nkR_{min}]\!]$ ($[\![.]\!]$ stands for the integer part), the size of the  $S$-matrix is then $N_{trunc}=2\Lambda+1$, with $\Lambda = \Lambda_{sc} + \Lambda_{ev}$.

\section{Root-search strategy}
\label{sect_rootsearch}
\begin{figure}[!hbt]
\centering
\includegraphics[width=0.4\linewidth,clip]{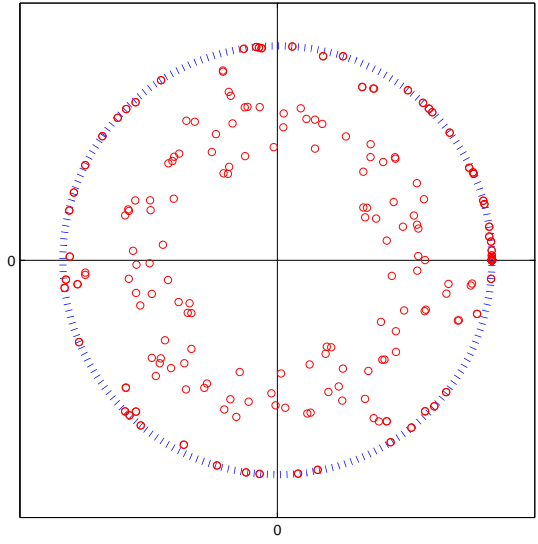} 
\caption{Distribution of scattering eigenvalues (red circles) in the complex plane for $nkR=106$, $\eps=0.12$, $n=2.65$. Blue dashed line is the unit circle $|z|=1$. Long-lived states have the modulus of the eigenvalue very close to unity, i.e.\ the eigenphase has only a small imaginary part $\eta$.}
\label{figEvaleps0.12}
\end{figure}

A typical run at $nkR_0=106$ for $\eps=0.1$ produces the eigenvalue distribution $\{\ex{i\varphi_k}\}$ plotted in Fig.~\ref{figEvaleps0.12} in the complex $z=\ex{i\varphi}$ plane. We will denote $\varphi = \theta + i\eta$, where $\theta$ and $\eta$ are real numbers, so that $|z| = \exp(-\eta)$. Note that $\Lambda_{sc}= [\![nkR_0 (1-\eps)]\!]= 93$ and we have included $\Lambda_{ev} = 55$ evanescent channels. The handling of numerical stability issues relating to the inclusion of such a large number of evanescent channels is outlined in section~(\ref{sect_numimp}).

Our first observation is that all the eigenvalues are strictly distributed within the unit circle $|z|=1$, i.e.\ $\im{\varphi} < 0$. This is because of the restriction of solutions to outgoing waves only. Furthermore, there is an accumulation of eigenvalues on the boundary of the circle, particularly at  $\theta=2\pi^+$. As we have established, an eigenvalue for which  $\varphi^{(l)}(k) = 2\pi$ within a given numerical precision yields a quantized mode of the resonator. However, we should resist the temptation to simply take all the scattering eigenstates whose eigenphases are $\varphi \approx 2\pi$ to be quantized. As was pointed out in Ref.\cite{DietzEPSU95} in the case of a closed system, there is an accumulation of scattering eigenphases at $\varphi  \approx 2\pi^+$, which {\em do not} correspond to proper physical eigenmodes of the resonator. These are modes, which are primarily composed of evanescent channels, and can easily be distinguished from  regular modes, because of their lack of k-dependence, as we shall see below.

\subsection{Zero deformation-Case of the rotationally symmetric dielectric}

A lot can be learned by way of a simple example. We will consider a case where we know the exact solutions, namely the dielectric circle. The exact eigenstates of the scattering matrix for the circle can be given a precise physical meaning in terms of classical processes in the short wavelength limit. They correspond to motion with a conserved angular momentum, or in terms of our notation in section~(\ref{sect_refbilliard}), a given impact angle $\sin\chi$ on the dielectric interface. The resulting scattering matrix is diagonal in the angular momentum representation. This signifies the fact that a ``channel" with a given $m$ upon encountering the boundary will be scattered to the same channel $m$, corresponding to specular reflection. The scattering matrix can be written as
\be
[\mcal{S} (k)]_{mm^{\prime}} = -\delta_{mm^{\prime}}
\frac{\beshp{m}{nkR}}{\beshm{m}{nkR}} f_m(k)
\label{Smatdicircle}
\ee
where the function $f_m(k)$ is given by the following expression
\be
f_m(k)= \left[ 1-n\frac{\mbox{H}_m^{+ \prime}(nkR)}{\beshp{m}{nkR}}
\frac{\beshp{m}{kR}}{\mbox{H}_m^{+ \prime}(kR)} \right]
\times \left[ 1-n\frac{\mbox{H}_m^{- \prime}(nkR)}{\beshm{m}{nkR}}
\frac{\beshp{m}{kR}}{\mbox{H}_m^{+ \prime}(kR)}\right]^{-1}
\ee
This form in terms of the particular ratios of Hankel functions will help us simplify the expressions considerably in the asymptotic limit $nkR \ra \infty$. Notice that when $f_m(k)=1$,
\be
[S_c(k)]_{mm^{\prime}} = -\frac{\beshm{m}{nkR}}{\beshp{m}{nkR}}
\delta_{mm^{\prime}}
\ee
is the {\em external} scattering matrix for the closed circular
cavity, which is unitary. Then our quantization condition
$[S_c(k)]_{mm^{\prime}}=1$ yields
\be
J_m(nkR)=0
\ee
which is the exact quantization condition for wavevectors $nk$ of a metallic cavity. Hence in the form Eq.~(\ref{Smatdicircle}), the corrections due to the openness of the system are lumped into the factor $f_m(k)$.

Let's first consider the diagonal elements of Eq.~(\ref{Smatdicircle}) for $m>nkR$. We will use the notation $\alpha=\cosh^{-1}(m/nkR)$, $\alpha^{\prime}=\cosh^{-1}(m/kR)$, $\beta=\cos^{-1}(m/nkR)$ and $\beta^{\prime}=\cos^{-1}(m/kR)$. Note that $\alpha^{\prime} > \alpha \gg 1$.  Using the large-order asymptotic representations for Bessel functions\cite{abramovitz} and with proper attention on exponentially small terms, it can be shown that\cite{tureci_thesis}
\be
[\mcal{S} (k)]_{mm} \sim 1 + i(1+2n) \ex{-2m\alpha}
\label{eqsmateva}
\ee
for $m \gg nkR$. As noted, these entries correspond to scattering of evanescent channels and result in eigenphases exponentially close to zero, $\varphi \sim (1+2n)\ex{-2m\alpha}$. Thus, the accumulation of eigenphases on the unit circle close to the quantization point $\varphi=2\pi$ in Fig.~\ref{figEvaleps0.12} can be linked to such extremely evanescent channels, which are not the physical modes of the cavity. These modes can be interpreted as {\em creeping waves}, which are evanescent modes which cling to the surface of the resonator\cite{nussenzweig_book}. Note that the number of such scattering eigenstates depends strongly on our choice of $\Lambda_{ev}$ in our numerical implementation.

Next, we will look at the {\em internally reflected} channels. These are obtained for the entries $kR<m<nkR$. The asymptotic form of the corresponding matrix elements are\cite{tureci_thesis}
\be
[\mcal{S} (k)]_{m} \sim \ex{i\Theta} \left[ 1-i\left(2n\sin\beta
\ex{-\alpha'} - i \frac{\ex{-2m\alpha'}}{n\sin\beta} \right) \right]
\ee
where $\Theta$, which is identical to the closed-circle eigenphase, is real and given by
\be
\Theta(k) = -2m(\beta -\tan\beta) - \frac{\pi}{2}
\ee
These channels yield eigenvalues which accumulate exponentially close to the unit circle $|z|=1$, but unlike the evanescent modes Eq.~(\ref{eqsmateva}), with arbitrary phases. Note that the exponentially small difference from $|z|=1$ represents the evanescent leakage which vanishes in the short wavelength limit.

\begin{figure}[hbt]
\psfrag{L}{$L$}
\psfrag{2beta}{$2\beta$}
\psfrag{chi}{$\chi$}
\centering
\includegraphics[width=5cm]{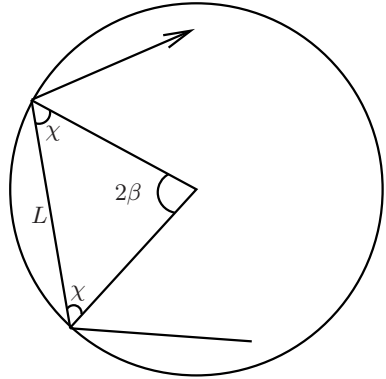}
\caption{Geometric representation of the angle $\beta$; the velocity of the eigenvalues in the complex plane is $2 \sin \beta$, which is also the chord-length of the corresponding ray. We note that for the diametral two-bounce orbit the speed is maximal (corresponding to the minimum free spectral range) while for whispering gallery modes the chord length is minimal and the free spectral range is the largest. }
\label{figbeta}
\end{figure}

It's possible to assign a velocity to these eigenphases in $k$-space:
\be
\der{}{\Theta}{(nkR)} = 2\sin\beta + O\left(\frac{1}{nkR}\right)>0
\label{quantspeed1}
\ee
A useful observation at this point is that this velocity is twice the cosine of the conserved ray impact angle $\chi=\pi/2 - \beta$ in the circular billiard corresponding to the motion with angular momentum $m$ (see Fig.~\ref{figbeta}).

The picture this entails is the following: When we slowly increase $k$, the individual eigenphases move with an approximately constant but mode-dependent speed given by Eq.~(\ref{quantspeed1}) counter-clockwise around the unit circle. Each time one of the eigenphases passes through $\varphi=2\pi$, the quantization condition is fulfilled and the resulting eigenvector is a quantized mode of the resonator. Hence, the eigenvectors of $\mcal{S} (k)$ can be assigned a physical meaning and identity even when $k$ is not tuned to resonance $\varphi(k)=2\pi$. In the present case, they correspond to totally internal reflected whispering gallery modes.

Last, we investigate the classically open channels, which corresponds to rays which are {\em refracted} out. In this regime $m<kR$ and
\be
[\mcal{S} (k)]_{m} \sim
\frac{\sin\beta'-n\sin\beta}{\sin\beta'+n\sin\beta} \, \ex{i\Theta}
\ee
Note that the algebraic prefactor is the Fresnel reflection factor for a ray coming in at an angle $\chi_i = \frac{\pi}{2} - \beta$. Thus, the proximity of the scattering eigenphase to the unit-circle is a measure of the lifetime. The smaller the radius of the eigenphase, the smaller is the associated lifetime. As we change $k$, the variation of the eigenphase of a given solution will be dominated by the phase-factor $\ex{i\Theta}$. The path to quantization goes thus by first increasing $\re{k}$ until $\re{\Theta}=2\pi$, and then adding a small imaginary part $i\Delta k$ so that
\be
|\ex{-i\Theta(k+i\Delta k)}| =
\frac{\sin\beta'-n\sin\beta}{\sin\beta'+n\sin\beta}
\ee
driving the eigenphase right to the quantization point. From this condition, we can extract an approximate value for the imaginary part of the quasi-bound mode which will result:
\be
\im{nkR} = -\frac{1}{2\sin\beta} \left| \log \left[
\frac{\sin\beta'-n\sin\beta}{\sin\beta'+n\sin\beta}
\right] \right|
\ee
This is precisely the lifetime of refractive WG modes due to Fresnel scattering, which can be obtained using different methods\cite{noeckel_thesis,tureci_thesis}.

The crucial point here is that these statements are only valid for an interval of the order of a mean-level spacing, so that $\beta$ is approximately constant
\be
\der{}{\beta}{(nkR)} = O\left(\frac{1}{nkR}\right)
\label{eqbetankr}
\ee
Furthermore, the assumption that $\im{nkR} \ll \re{nkR}$ is also implicit in these derivations. These procedures have to be implemented carefully because of the Stokes phenomenon\cite{orszag_book,bleistein_book} in the asymptotic expansion  of the Hankel functions with complex argument. However, as long as the latter condition is satisfied, these estimates are valid.

\subsection{Deformed dielectric resonators}

In light of our findings for the undeformed case, it is possible to develop a powerful search strategy for the general, deformed case. The reason behind our ability to ``track" the scattering eigenphases through quantization in the case of the circular resonator was the fact that the angular momentum channels didn't mix when we changed $k$, owing to the diagonality of the scattering matrix over all $k$ i.e.\ there we had a good label $m$ which was conserved. This will not be the case when we deform the resonator. For small deformations, the internal scattering matrix $\mcal{S} (k)$ will remain approximately diagonal, with fluctuations due to inter-channel scattering. The resulting eigenstates will show a broadening in their angular momentum distributions. In that case, one can still define an average phase velocity given by
\be
\der{}{\bar{\Theta}}{(nk\bar{R})} = 2\sin\bar{\beta}
\label{phasespeedav}
\ee
defined by the average angular momentum  $\bar{m}$
\be
\bar{\beta}=\cos^{-1}\frac{\bar{m}}{nkR}\qquad\bar{m} = \frac{1}{2\Lambda+1} \sum_{-\Lambda}^{\Lambda} m |\alpha_m|^2
\ee
At first sight, there is no reason for such a solution to persist over a given interval $\Delta k$. Following Ref.\cite{frischat97}, we suggest that the scattering eigenvectors have an identity beyond a given $k$-value, and more importantly, that the resonances, the quantized modes, have an identity even when they don't fully satisfy the boundary conditions. We can quantify this statement by defining a simple scalar product between eigenvectors of the internal S-matrix at different $k$:
\be
\braket{\alpha(k)}{\alpha(k+\Delta k)} = \sum_m
\alpha_m(k)\alpha_m^*(k+\Delta k)
\label{dotprodalpha}
\ee

\begin{figure}[!hbt]
\psfrag{overlap}{$\braket{\alpha(k_0)}{\alpha(k)}$}
\centering
\includegraphics[width=0.7\linewidth,clip]{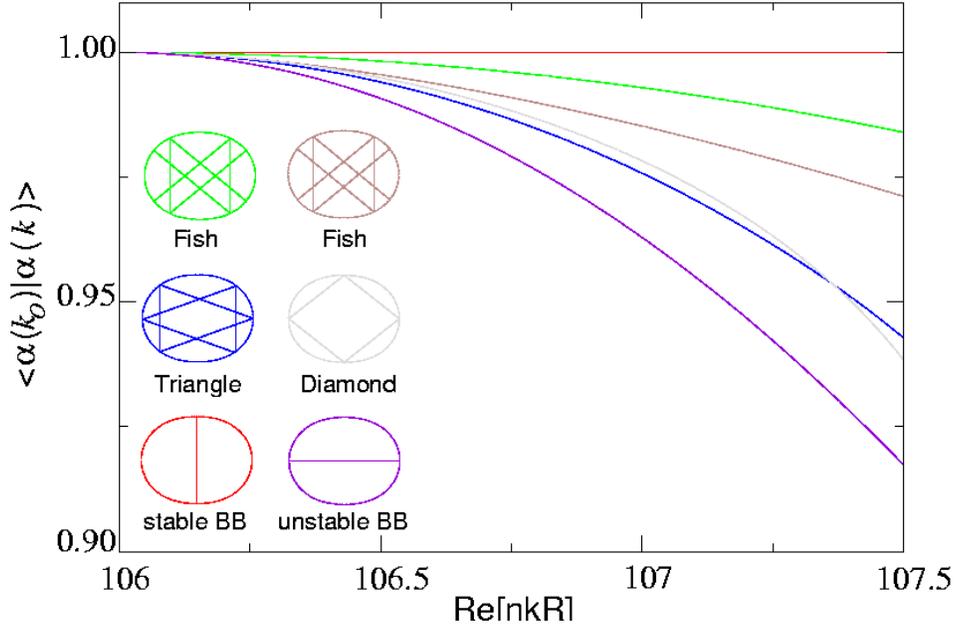}
\caption{The overlap calculated for a set of states in the interval $nkR=106-107.5$, for $\eps=0.12$ and $n=2.65$. The associated classical structures are found from the Husimi projections of the respective states (see Fig.~\ref{figcrossing}).  The schematics identify the ray orbits with which they are associated (see discussion in text).  At this deformation the short two-bounce orbit is stable, the long one unstable; the diamond orbit is stable and the fish and triangle orbits are unstable.}
\label{figatracegood}
\end{figure}

Then our claim is tantamount to the adiabaticity of $\braket{\alpha(k)}{\alpha(k+\Delta k)}$. The reason this is possible lies in the subtle correlations among the matrix elements induced by the underlying classical motion in the short wavelength limit. We have already emphasized the connection between the scattering matrix in the short wavelength limit and the classical SOS map. As long as there are invariant curves in the SOS, which we have seen is guaranteed by the KAM scenario for near-integrable deformations in section~(\ref{sect_refbilliard}), there will be eigenstates of the scattering matrix which will display the aforementioned adiabatic behavior.

In Fig.~\ref{figatracegood} we trace the overlap Eq.~(\ref{dotprodalpha}) of a set of eigenvectors in an interval of the order of a mean-level spacing. First, a diagonalization of $\mcal{S} (k)$ is performed at a $k_0$, the eigenvectors determined, and then further diagonalizations are performed at regular intervalls $k=k_0+j\Delta k$, where $n\Delta kR = 0.03$. At each step, there is in general a single state having markedly higher overlap with the respective {\em original state} at $k_0$ than the others and that value is plotted. The result shows that an adiabatic identity can be in fact defined for certain states. This procedure allows the tracking of majority of the states, as long as the deformation is not too large. In fact, it's possible to show that
\be
\braket{\alpha(k)}{\alpha(k+j\Delta k)} = 1 + jn\Delta kR \cdot
O(\frac{1}{nkR})
\ee

\begin{figure}[!hbt]
\psfrag{overlap}{$\scriptstyle\braket{\alpha_i(k)}{\alpha_0}$}
\psfrag{bra0}{$\scriptstyle\bra{\alpha_0}$}
\psfrag{bra1}{$\scriptstyle\bra{\alpha_1}$}
\centering
\includegraphics[width=\linewidth,clip]{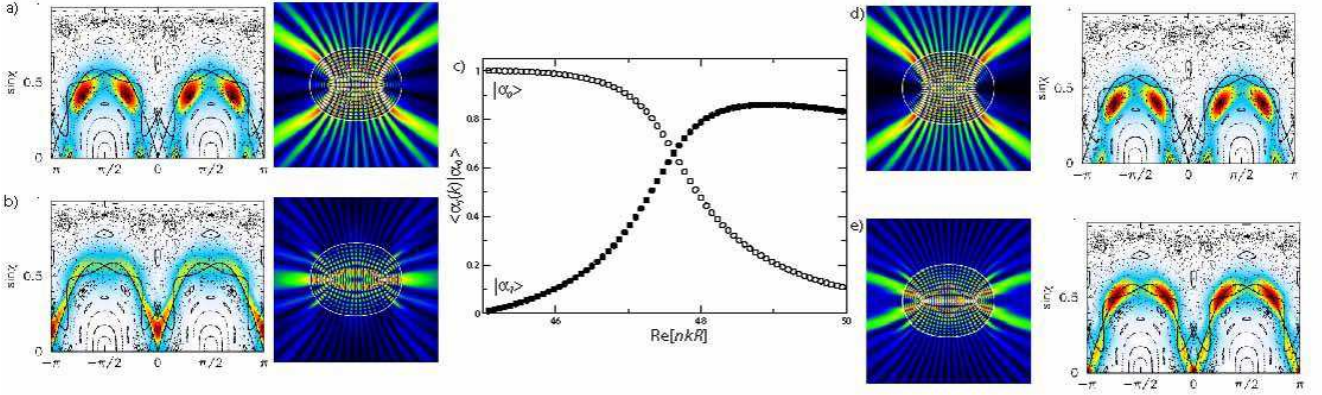}
\caption{Two eigenvectors are traced by the criterion that the overlap is largest in two {\em consecutive} iterations. The figure shows the overlap of the two sets of states resulting with respect to one of the {\em initial} states, $\bra{\alpha_0}$.  Away from the avoided crossing the states have distinct classical meaning as discussed in the text; they exchange ``identity" at the avoided crossing. Both the real-space electric field intensities are shown (false color scale) and the Husimi-SOS projections of the states before and after the avoided crossing.}
\label{figcrossing}
\end{figure}

At this point it may be helpful to clarify what we mean by the ``identity" of a state in the chaotic case in which the state is not associated with a stable periodic orbit or a family of quasi-periodic orbits.  In Fig.~\ref{figcrossing} we show both a real-space solution for the electric field of a TM resonance and its projection onto the surface of section using the Husimi-SOS projection technique defined in section~(\ref{sect_husimi}).  This method allows one to associate any solution, even a non-quantized one, with a region in the SOS and hence with an approximate ray-dynamical (classical) meaning. Furthermore, at the values of $nkR$ at which we work, often these chaotic states are associated with unstable periodic orbits or their unstable manifolds (this is the case in Figs.~\ref{figatracegood} and~\ref{figcrossing}); in Fig.~\ref{figcrossing} the top states are associated with the unstable fish orbit and the bottom states are associated with the unstable manifolds of the unstable two-bounce orbit along the major axis of the resonator. States localized on unstable periodic orbits have been termed ``scars" in the quantum chaos literature and will be discussed further in section~(\ref{sect_qbm}).

It turns out that one can extend this strategy to higher deformations, where the SOS displays large chaotic components, with proper attention to eigenstates which have an appreciable overlap with chaotic regions. A typical scenario which is encountered is the {\em avoided crossing} of two scattering eigenvectors. This is captured in Fig.~\ref{figcrossing}, where two eigenvectors are traced over a mean-level spacing. Originally, the two states are well-distinguished; they have approximately zero overlap with each other. They have different classical meaning as well as shown by their Husimi projections on the SOS (see section~(\ref{sect_husimi}) for definition). One state is associated with the border of the stable bouncing ball region of the SOS and has no intensity near $\phi = 0,\pi$; the other is concentrated in the separatrix region associated with the unstable period two orbit along the major axis (we have plotted its unstable manifold for reference). At the crossing they perturb each other strongly, and an approximate superposition state results. However, if we continue changing $k$, the states emerging from the avoided crossing will still have a pronounced overlap with the states before the crossing. Notice that the overlaps are calculated with reference to one of the original states $\bra{\alpha_0}$. This example represents a case where a numerical tracing algorithm has to be properly conditioned.

\begin{figure}[!t]
\centering
\psfrag{dphi}{$d\theta / d(\re{nkR})$}
\psfrag{detaeta0}{ $d\eta / d(\im{nkR})$}
\includegraphics[width=\linewidth]{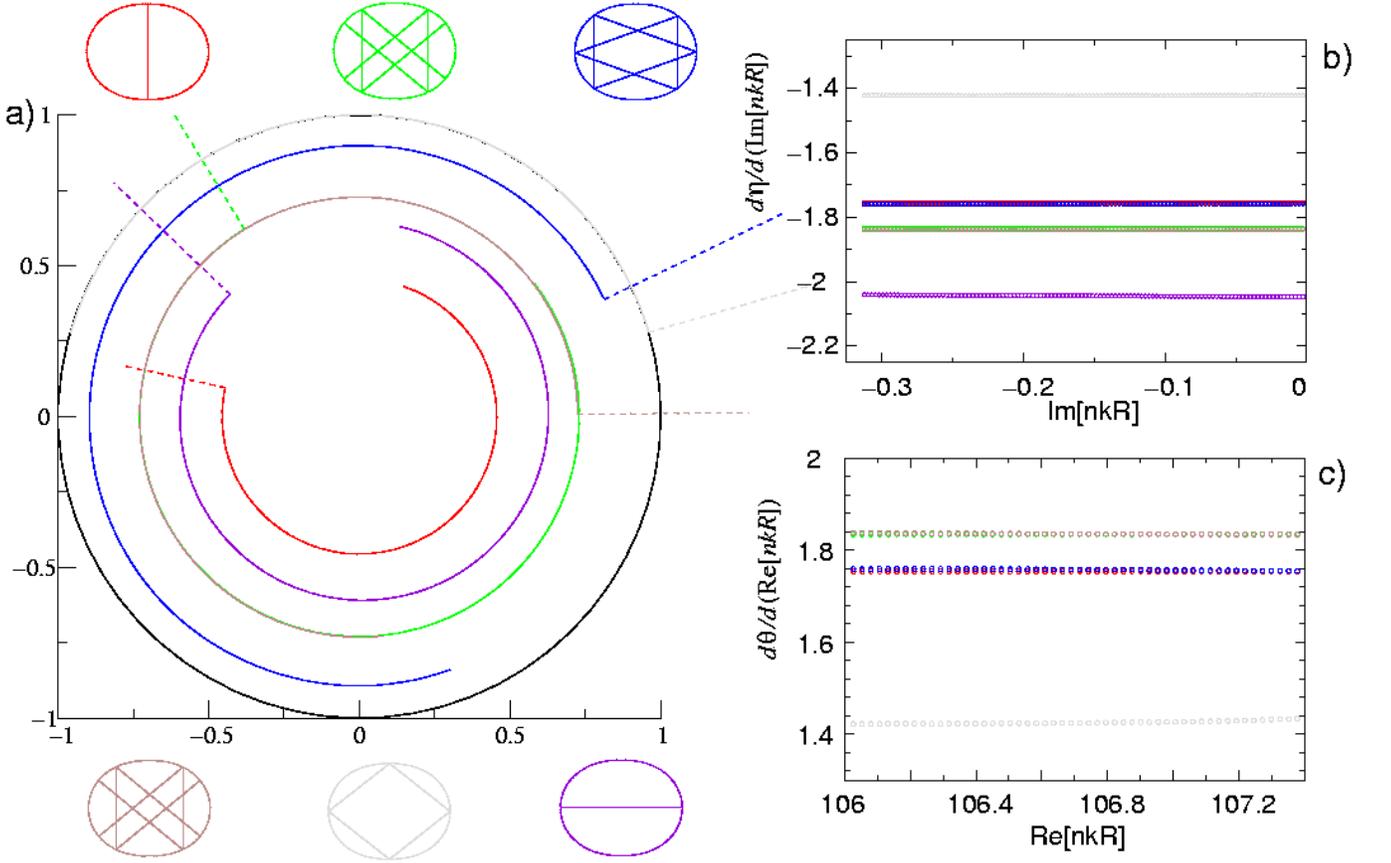}
\caption{Several representative eigenvalues $z$ (corresponding to states associated with the ray orbits above and below via color code) traced in the complex plane as one changes the real and imaginary values of $k$. First $\re{k}$ is varied resulting in the circular arcs of fixed radius ($\im{z}$); subsequently $\im{k}$ is varied resulting in a radial motion and fixed $\re{k}$. On the right hand side we show the constancy of the derivative of the phase angle $\theta$ with respect to $\re{nkR}$ and of the derivative of the logarithm of the radius $\eta(k)$ with respect to $\im{nkR}$, implying constant speed of the eigenphases as a function of $k$ in the complex plane. The simulations are performed at $\eps =0.12$ quadrupolar deformation, $n=2.65$.}
\label{figphasescancircle}
\end{figure}

After having established that we can assign an identity to the scattering eigenvectors as $k$ varies, we next investigate how precisely the corresponding eigenvalues move within the complex unit circle as we vary $k$, both through real and imaginary values. Fig.~\ref{figphasescancircle}a) shows such a tracing of several representative states. First, the initial eigenvalues are followed while varying the real part of $k$; each of the eigenvalues follow approximately a circular trajectory, followed by a pure imaginary change in $k$ resulting in the eigenvalues following an almost precisely radial path. We write the radius of the complex eigenvalue as $|\ex{i\varphi(k)}|=\ex{\eta(k)}$ and call $\theta$ the angle in the complex plane for the eigenvalue $\ex{i\theta(k)}$.

This simple behavior can be understood from the fact that the classical channels (of angular momentum in our case) in the expansion preserve their identity over a mean level spacing, and the weight of these channels embodied in the expansion coefficients $\alpha_m$ change only $O\left(\frac{1}{nkR}\right)$. In conclusion, the radial and angular speeds of the eigenvalue are approximately ``decoupled". This speed is to high accuracy constant for the eigenphases, i.e. for the log of the eigenvalues as shown in Fig.~\ref{figphasescancircle}b, c.

We have developed an efficient numerical algorithm to determine the quasi-normal modes of an smoothly deformed dielectric resonator based on all of these observations:
\begin{enumerate}
\item A diagonalization of $\mcal{S} (k)$ is performed at a given $k$, and $N_{trunc}$ eigenphases and eigenvectors are determined, denoted by $\bra{\alpha^{(i)}_0}$, $i=1,\ldots N_{trunc}$; $\braket{m}{\alpha^{(i)}_0} = \alpha_m^{(i)}$.
\item A second diagonalization is performed at $k+\Delta k$, where $\Delta k$ is a small complex number so that $|\Delta k| \ll k$.
\item Approximate radial and angular eigenphase speeds are determined.
\item Assuming the constancy of the individual speeds, an approximate quantization wavevector $k_q^{(i)}$ is determined for each of the initial eigenvectors.
\item Finally, the quasi-bound modes are constructed by
\end{enumerate}
\be
\psi^{(i)}_q(r,\phi) = \sum_{m=-\Lambda}^{\Lambda} \alpha_m^{(i)}
\besj{m}{nk_q^{(i)}r} \ex{im\phi}\label{eqqbm13}
\ee
Due to the small change in $\{ \alpha_m^{(i)} \}$ with $k$, we simply use their non-quantized values in the expansion with the extrapolated $k$-value $k_q^{(i)}$. We have checked that it's important to use $k_q^{(i)}$ instead of the original $k$.

In the ideal case this means that $N_{trunc}  \sim nkR$ quasi-bound modes are found in only two diagonalizations. In practice, this ideal limit is not fully attained. But, depending on the deformation and the value of $nkR$, a large fraction of the quasi-bound modes can be calculated approximately in this manner. Table Fig.~\ref{tabletracing} shows a typical run and the quality of the results compared to ``exact" solutions.  For increased numerical stability we have found it convenient to use the same algorithm in the context of solving the equivalent generalized eigenvalue problem for the system; this is discussed briefly in the Appendix~\ref{sect_numimp}.

\begin{table}
\caption{A typical run at $kR_0=40$, $\eps=0.12$ and $n=2.65$. The first column represents the predicted value assuming a constant speed as determined from two successive diagonalizations separated by $\Delta kR = 10^{-4}+i10^{-4}$. The second column is the error of this prediction, obtained by a full root search, measured by the distance in the complex plane between the eigenphase and the quantization point. The last two columns are the overlaps of the original eigenvectors (internal and external) with the actual quantized ones.}
\begin{tabular}{|c|c|c|c|}
\hline
         $kR_q$        &     $|\ex{i\varphi}-1|$ &
$\braket{\alpha_q}{\alpha_0}$      &    $\braket{\gamma_q}{\gamma_0}$
\\
\hline
$  40.530139923096 -i0.128341300297e-03$ &
0.1049356E-01  &  0.935212779    & 0.815777069  \\
$  40.354640960693 -i0.346842617728e-02$ &
0.2512625E-01  &  0.800668216    & 0.888131129  \\
$  40.362663269043 -i0.262046288699e-01$ &
0.3483059E-01  &  0.875476530    & 0.959173141  \\
$  40.597846984863 -i0.617872737348e-02$ &
0.4986330E-01  &  0.885584168    & 0.902043947  \\
$  40.760002136230 -i0.517168489750e-03$ &
0.2216994E-02  &  0.900596963    & 0.598602624  \\
$  39.372772216797 -i0.782183464617e-02$ &
0.3692186E-01  &  0.670832741    & 0.857146071  \\
$  39.384689331055 -i0.253524887376e-02$ &
0.1560470E-01  &  0.644713713    & 0.768504668  \\
$  39.524833679199 -i0.416035996750e-03$ &
0.6675268E-02  &  0.918124783    & 0.571120689  \\
$  40.427906036377 -i0.654014274478e-01$ &
0.2790542E-02  &  0.989788008    & 0.989591927  \\
$  40.367130279541 -i0.640191137791e-01$ &
0.2049567E-01  &  0.949528775    & 0.983221172  \\
$  40.508068084717 -i0.814560204744e-01$ &
0.1035770E-02  &  0.979746925    &0.983924594  \\
$  40.537075042725 -i0.717425644398e-01$ &
0.6688557E-01  &  0.910022901    & 0.939244195  \\
$  40.627620697021 -i0.913884192705e-01$ &
0.8095847E-02  &  0.858722180    & 0.846152346  \\
$  39.421646118164 -i0.850722268224e-01$ &
0.5916540E-01  &  0.906494113    & 0.943557886  \\
$  39.419769287109 -i0.566106282349e-04$ &
0.1063111E-01  &  0.939688864    & 0.531911062  \\
$  39.733528137207 -i0.175021495670e-01$ &
    0.3724663E-01 &  0.783724495    & 0.988899106  \\
$  39.563480377197 -i0.650179386139e-01$ &
    0.1242726E-02 &  0.990091490    & 0.989958495  \\
$  39.685890197754 -i0.770573019981e-01$ &
    0.1809454E-01 &  0.936621212    & 0.951369186  \\
$  39.650619506836 -i0.790241658688e-01$ &
    0.1439820E-01 &  0.934294618    & 0.935693384  \\
$  39.369716644287 -i0.169137448072e+00$ &
    0.3348814E-03 &  0.999949714    & 0.999923966  \\
$  40.361137390137 -i0.111876547337e+00$ &
    0.3443463E-02 &  0.998603169    & 0.996169638  \\
$  40.389163970947 -i0.635751348455e-03$ &
    0.3869480E-02 &  0.902893323    & 0.686779751  \\
$  40.273880004883 -i0.709924623370e-01$ &
    0.1889875E-02 &  0.987924322    & 0.995055099  \\
$  40.207023620605 -i0.329816946760e-02$ &
    0.5157299E-01 &  0.870880666    & 0.955577617  \\
$  40.213146209717 -i0.278946310282e-01$ &
    0.4355001E-01 &  0.850281901    & 0.996592453  \\
$  39.904693603516 -i0.224154405296e-01$ &
    0.2102797E-02 &  0.998176457    & 0.998674123  \\
$  40.172363281250 -i0.132210448384e+00$ &
    0.3454306E-03 &  0.999768065    & 0.999390566  \\
$  40.137271881104 -i0.873498693109e-01$ &
    0.6845250E-03 &  0.997220268    & 0.997824221  \\
$  40.025264739990 -i0.162262007594e+00$ &
    0.1378969E-03 &  0.999992333    & 0.999984896  \\
$  40.064556121826 -i0.597218498588e-01$ &
    0.1994631E-03 &  0.998558338    & 0.999194260  \\
$  40.058116912842 -i0.746092235204e-03$ &
    0.6044292E-02 &  0.948958619    & 0.983942653  \\
$  39.997570037842 -i0.659425705671e-01$ &
    0.1710776E-02 &  0.999267906    & 0.999318229  \\
$  39.925022125244 -i0.229551533266e-04$ &
    0.5157831E-04 &  0.999120149    & 0.934566350\\
\hline
\end{tabular}\label{tabletracing}
\end{table}

The implementation of the algorithm can be adapted to the particular result of interest. In fact, when the quantization of a single state is desired, a more exact eigenphase quantization can be performed by multiple-scans with update of the speeds, quite like Newton's root-search method.

As already hinted, the method is most powerful when applied to calculate classically meaningful quantities (such as the Husimi projection, see section~(\ref{sect_husimi}) below) for which finding the quantized $k$ values is not important. This is most evident in considering the physical observables which are unique to the open systems, the farfield emission patterns and the boundary image fields. The calculation of these observables will be described below.

On the other hand, it has to be pointed out that an attempt to plot an internal solution away from quantization is not meaningful because of the nature of the Hankel function basis. The internal field would
be
\be
\psi_1(r,\phi) = \sum_m \alpha_m (\beshp{m}{nkr} +
\ex{i\varphi}\beshm{m}{nkr}) \ex{im\phi}
\ee
The existence of the factor $\ex{i\varphi}$ exposes the $z^{-m}$
singularity at the origin due to the Neumann components
$\mbox{N}_m(z) = i(\beshm{m}{z} - \beshm{p}{z})$. That is one reason
that we must use $k_q$ in Eq.~(\ref{eqqbm13}). This limitation is an artifact of
the particular basis used, and does not arise for example for the
case of an expansion in cartesian modes.

\section{The Husimi Projection technique for optical dielectric resonators}
\label{sect_husimi}
In this section, we will describe the Husimi Projection technique, which allows us to relate a given mode to the phase space structures in the SOS. Just as for quantum wavefunctions, for these two-dimensional electromagnetic fields we can represent the solutions in real space (the solutions we have been calculating) or, by Fourier transforming them, in momentum space. However we are interested in representing the solutions in the phase-space of the problem so that we can understand their ray-dynamical meaning, and ultimately in projecting such phase space densities onto the SOS which is our standard interpretive tool. Just as in quantum mechanics, we cannot have full information about real-space and momentum space at the same time due to the analog of the uncertainty principle, which here is reflected by the property of Fourier transforms:
\be
\Delta x \cdot \Delta p  \ge \frac{1}{2k}
\label{heisenberg}
\ee
where $\Delta x$ and $\Delta p$ are the widths of $\psi(\bm{x})$ and $\tilde{\psi}(p)$ in real and momentum space. Thus our goal is to take the solution $\psi (\bm{x})$ and associate with it a momentum content in some region around each point $\bm{x}$, recognizing that our resolution in real space is limited by the uncertainty relation. One familiar method for doing this in quantum mechanics is the Wigner distribution function\cite{wigner32}, which preserves exactly certain moments of the wavefunction, but has the interpretive problem that it can be negative in some regions of phase space and the practical problem that it is typically subject to rapid oscillations. An alternative approach, which is often more useful, was introduced by Husimi\cite{husimi40} and can be thought of as a gaussian smoothing of the Wigner distribution. In our context however one can think of the Husimi projection as a ``windowed" two-dimensional Fourier transform, which is designed to respect the uncertainty relation. This is equivalent to projecting on coherent states $\bra{z}=\bra{\bar{\bm{p}},\bar{\bm{x}}}$ which are optimally localized in both momentum {\em and} configuration space, in the sense that the projection saturates the inequality Eq.~(\ref{heisenberg}) at its lower bound
\be
\Delta x = \frac{\sigma_0}{\sqrt{2k}}=\frac{\eta}{\sqrt{2}} \qquad
\Delta p = \frac{1}{\sqrt{2k}\sigma_0}
\label{etabound}
\ee
{\em and} equal resolution in both spaces is ensured through the choice of $\sigma_0$. This is a free parameter of the method, and for best results it has to be carefully determined based on the domains of variation of the conjugate pair $(\bm{p},\bm{x})$. The real-space representation of a coherent states is given by
\be
Z_{\bar{\bm{x}}\bar{\bm{p}}}(\bm{x}) =
\left(\frac{1}{\pi\eta^2}\right)^{1/4}
\exp(ik\bar{\bm{p}}\cdot\bm{x})
\exp\left(-\frac{1}{2\eta^2}|\bm{x}-\bar{\bm{x}}|^2\right)
\ee
where the prefactor ensures normalization. Note that the first exponential factor determines the selection of the momentum $\bar{\bm{p}}$ (normalized to unity), and the second exponential factor restricts the probe to an isotropic window of size $\Delta x = \eta/\sqrt{2}$ around $\bar{\bm{x}}$ in the configuration space. The Fourier transform is again a localized Gaussian, given by
\be
\tilde{Z}_{\bar{\bm{x}}\bar{\bm{p}}}(\bm{p}) =
\left(\frac{\eta^2}{\pi}\right)^{1/4}
\exp(-ik(\bm{p}-\bar{\bm{p}})\cdot\bar{\bm{x}})
\exp\left(-\frac{\eta^2}{2}|\bm{p}-\bar{\bm{p}}|^2\right)
\ee
Note that {\em both} scales $\Delta x$ and $\Delta p$ are sharpened as $k\ra \infty$. We can now construct Husimi distribution by projecting on these states
\be
\rho_\psi(\bar{\bm{x}},\bar{\bm{p}}) = |\braket{z}{\psi}|^2 = \left|
\int d^2\bm{x} Z_{\bar{\bm{x}}\bar{\bm{p}}}^*(\bm{x}) \psi(\bm{x}) \right|^2
\ee
Note that this distribution, unlike the Wigner distribution, is positive definite in the phase space $(\bar{\bm{x}},\bar{\bm{p}})$.

As defined the Husimi distribution is on a four-dimensional phase space of the billiard (restricted to a three-dimensional shell because the momentum is normalized to unity).  One can visualize this distribution then as a vector field on a grid of size $\eta^2/2$ in real space\cite{heller_leshouches89}; however based on our earlier discussion we find it more illuminating to define a projection of the Husimi distribution onto the surface of section of the billiard.

For billiard systems, this idea was first carried out in Ref.\cite{crespiPC93}, but a different section was used than we are using here. We instead follow an approach similar to the one described in Ref.\cite{hackenbroichN97}. The problem is that the boundary of the billiard is not a constant coordinate surface of some convenient coordinate system. We therefore use cylindrical coordinates to define our Husimi-SOS projection and then use the classical map to map our result onto the boundary.

The coherent states in cylindrical coordinates take the form:
\be
Z_{\bar{r}\bar{\phi}\bar{p}_r\bar{p}_\phi}(r,\phi) =
Z_{\bar{r}\bar{p}_r}(r)Z_{\bar{\phi}\bar{p}_\phi} (\phi)
\ee
where
\beq
Z_{\bar{r}\bar{p}_r}(r) & = & \left(\frac{1}{\pi\eta^2}\right)^{1/4}
\exp\left[i\bar{p}_r r\right]\exp\left[-\frac{1}{2\eta^2} (r-\bar{r})2\right] \\
Z_{\bar{\phi}\bar{p}_\phi} (\phi) & = &
\left(\frac{1}{\pi\sigma^2}\right)^{1/4} \sum_{l=-\infty}^{\infty}
\exp\left[i\bar{p}_{\phi} (\phi-2\pi l)\right] \exp\left[-\frac{1}{2\sigma^2}
(\phi-\bar{\phi}-2\pi l)^2\right]
\eeq
The sum on $l$ is necessary to insure periodicity in the $\phi$
variable. We define the projection
of the full four-dimensional Husimi function onto the SOS at constant
radius $r=R_c$ by
\be
H(\phi,p_\phi) = \lim_{\eta\ra 0} \int_0^{\infty} d\bar{p}_r\left|
\int_{-\pi}^{\pi} d\phi \int_0^{\infty} dr
Z^*_{R_c\bar{\phi}\bar{p}_r\bar{p}_\phi}(r,\phi) \psi(\phi,r)
\right|^2
\label{eqhusimiraw}
\ee
Note that the integration only extends over positive radial momenta to accord with the definition of the SOS which only counts trajectories which encounter the boundary in the outgoing direction. We have included in the definition Eq.~(\ref{eqhusimiraw}) the limit $\eta \rightarrow 0$ representing the short wavelength limit where $k \rightarrow \infty$.  In practice, due to Eqs.~(\ref{heisenberg}),~(\ref{etabound}),  for any given $k$ value $\eta$ is bounded below and is greater than the wavelength $\lambda$. Violation of this condition would give unphysical results. For example, a closed circular billiard would have $\psi (r=R_c) = 0$ and the integrand of Eq.~(\ref{eqhusimiraw}) would vanish if $\eta$ were taken to zero before $k \rightarrow \infty$ due to the concentration of the coherent state to a region less than a wavelength from the boundary; hence the Husimi-SOS would vanish. Letting $\eta \ra 0$ and $1/k \ra 0$ while $k\eta>1$, it can be shown that
\be
\lim_{\eta\ra 0} \int_0^{\infty} d\bar{p}_r\left| \int_{-\pi}^{\pi}
d\phi \int_0^{\infty} dr
Z^*_{R_c\bar{\phi}\bar{p}_r\bar{p}_\phi}(r,\phi) \psi(\phi,r)
\right|^2 \approx \left| \int_{-\pi}^{\pi} d\phi
Z^*_{\bar{\phi}\bar{p}_\phi}(\phi) \psi^{(+)}(R_c,\phi) \right|^2
\label{eqsosreduct}
\ee
where $\psi^{(+)}(R_c,\phi)$ contains only the wavefunction
components with the Hankel's functions of the first kind:
\be
\psi^{(+)}(R_c,\phi) = \sum_{m=-\infty}^{\infty} \alpha_m
\beshp{m}{nkR_c} \ex{im\phi}
\label{eqposhank}
\ee
The presence of Hankel's functions of only one kind is in accordance with the short wavelength interpretation that $\beshpm{m}{nkr}$ represents incoming and outgoing waves respectively. This interpretation is quickly obscured for components $m>nkR$, which however have vanishing weights in the expansion Eq.~(\ref{eqposhank}) through the physical considerations laid out in section~(\ref{sect_scatquant1}).

Using Eq.~(\ref{eqsosreduct}) and the short wavelength correspondence $p_\phi \leftrightarrow m = nkR_c \sin\chi$ in Eq.~(\ref{eqhusimiraw}) we obtain
\be
H_{\psi}(\bar{\phi},\bar{\sin\chi}) = \left(\frac{1}{\pi\sigma^2}\right)^{1/4}
\sum_{l=-\infty}^{\infty} \int_{-\pi}^{\pi} d\phi \exp\left[-inkR_c
\bar{\sin\chi} (\phi-2\pi l) \right]
\exp\left[
- \frac{1}{2\sigma^2}(\phi -\bar{\phi}-2\pi l)^2 \right] \Psi^{+}(\phi)
\label{husimi_dist}
\ee
Noting that the integrand in Eq.~(\ref{husimi_dist}) is $2\pi$-periodic, the integration limits can be extended to infinity and the resulting Gaussian integral over $\phi$ can be evaluated analytically, yielding
\be
H_{\psi}(\bar{\phi},\sin \bar{\chi}) = \left| \sum_{-\infty}^{\infty}
\alpha_m \beshp{m}{nkR_c} \ex{-inkR_c(\sin\chi- \sin\bar{\chi})
\bar{\phi}} \ex{-\frac{\sigma^2}{2} (\sin\chi- \sin\bar{\chi})^2}
\right|^2
\label{husimicirclesect}
\ee
For optimal resolution in both SOS coordinates, we choose $\sigma \sim 1/\sqrt{nkR_c}$ ($R_c$ can be chosen at any convenient value). Finally, in order to calculate the distribution on a section on the boundary $r=R(\phi)$ instead of the circle, we choose $R_c$ slightly outside the boundary, and extrapolate the Husimi distribution Eq.~(\ref{husimicirclesect}) to the boundary using the classical equations of motion; i.e.\ every pair $(\sin \bar{\chi},\bar {\phi})$ on the circle maps uniquely to a different pair $(\sin \chi, \phi)$ on the boundary.  We can regard this mapping as simply a change of variables so that the Husimi-SOS distribution on the boundary, $H_{\psi}(\phi,\sin \chi)$, satisfies:
\be
H_{\psi}\left(\phi(\bar{\phi},\sin\bar{\chi}),\sin\chi(\bar{\phi},\sin\bar{\chi})\right) = {H}_{\psi}\left(\bar{\phi},\sin\bar{\chi}\right)
\ee

\section{Farfield Distributions}
\label{sect_FF}
In typical micro-laser experiments there are two basic data-acquisition modes. In the {\em farfield} acquisition mode, the CCD camera is used without a lens and aperture as a simple photo-diode, and at each farfield angle $\theta$, the farfield emission intensity $I_{FF}(\theta)$ is recorded. The farfield emission pattern is one of the few clues we have as to the lasing mode of the resonator. Our algorithm is very efficient in calculating all possible farfields achievable with a given resonator shape and index of refraction. At a given $k$ which is chosen close to the lasing frequency $\omega=ck$, we solve the scattering problem without reference to any quantization condition. The farfields are computed from the external wavefunction Eq.~(\ref{extwf}) by using the large-argument asymptotic form of the Hankel function\cite{abramovitz}
\be
I(\phi) \propto \left| \sum_m \gamma_m
\ex{im\left(\phi-\frac{\pi}{2}\right)}  \right|^2,
\ee
where we have extracted all the quantities independent of $m$ and
$\phi$. The farfield intensity distributions computed in this way are
very well-behaved and insensitive to the k-value used, because the
strongly-varying Hankel functions drop out of  this quantity.

\begin{figure}[!hbt]
\centering
\psfrag{quantizedA}{\footnotesize quantized @ $kR=19.739 - 0.0624i$}
\psfrag{AAA}{\footnotesize $z= -0.4664 + 0.5692i$}
\psfrag{l2}{\footnotesize $z=+0.3407 - 0.6472i$}
\psfrag{quantizedB}{\footnotesize quantized @ $kR=21.021 - 0.0624i$}
\psfrag{real}{$\re{z}$}
\psfrag{imag}{$\im{z}$}
\psfrag{theta}{Far-field angle $\theta$}
\psfrag{intensity}{Intensity a.u.}
\includegraphics[width=13cm]{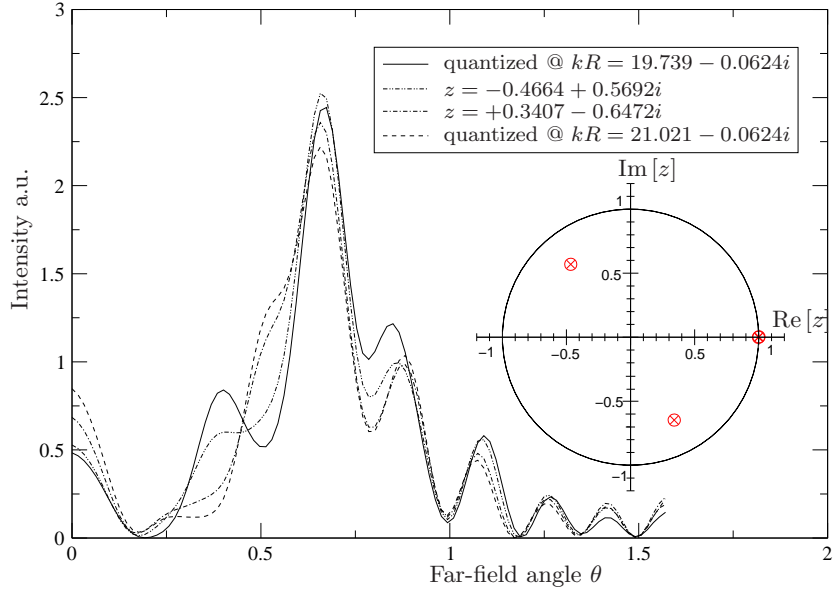}
\caption{Far-field emission intensity pattern for a ``fish" mode, quantized at $kR=19.7392-0.06240i$
(solid black line) and at $kR=21.0210-0.06240i$ (dashed line). The
(dash dot) and (dotted) line are the far-field of the unquantized
states in between with eigenvalues $z =-0.4664 +0.5692i$ and
$z =0.3407-0.6472i$.  Clearly the essential features of the emission patterns are calculable with non-quantized modes.  In the inset we show the eigenvalues
$z$ in the unit circle.}
\label{figFFFish}
\end{figure}

As seen in Fig.~\ref{figFFFish}, the farfields computed are virtually
identical to those obtained from the quantized modes as $k$ is varied
over two level-spacings.  Using this simplification the method here
can be used to evaluate all possible emission patterns for a wide
range of dielectric resonators very rapidly.

\begin{figure}[!hbt]
\centering
\includegraphics[width=0.7\linewidth]{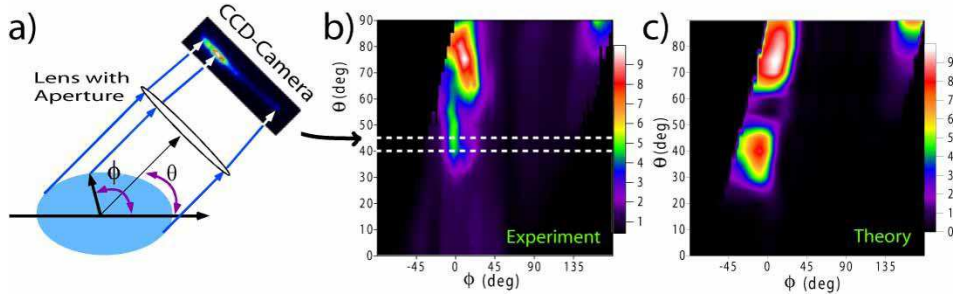}
\caption{a) Experimental setup for measuring simultaneously far-field
intensity patterns and boundary images as implemented in ref.\cite{rex02}; the lens and aperture are in the far-field and the schematic is not to scale. b) experimental data on lasing emission from the same reference.  c) Simulation of the experimental set-up based on calculated modes of the corresponding quadrupolar ARC dielectric resonator.}
\label{figsetupnathan}
\end{figure}

Several recent experiments have studied dielectric micro-lasers using
an imaging technique for data acquisition\cite{rex01,rex02,schwefel03,grace03}; the CCD camera records a magnified image of the intensity profile on the sidewall
viewed from angle $\theta$ in the farfield (see Fig.~\ref{figsetupnathan}). The pixels then can be
mapped to sidewall angle $\phi$ via the solution of a simple
transcendental equation. This yields a two-dimensional plot, called
the {\em boundary image-field}, where a given data point
$I(\phi,\theta)$ denotes the intensity emitted from sidewall position
$\phi$ towards the farfield angle $\theta$. The latter can easily be
converted to an incidence angle $\sin\chi$, using Snell's law and
basic trigonometry. Hence what is recorded is actually a phase space
plot of the emitted radiation.

The aperture has an important role of defining a window in the direction space ($\Delta\sin\chi$), so that a given pixel on the camera can be identified up to a diffraction limited resolution with a pair ($\phi,\sin\chi$). Mathematically, the effect of the lens-aperture combination is equivalent to a windowed Fourier transform of the incident field on the lens\cite{goodman_book}; thus it is simply connected to the Husimi distributions we have just discussed\cite{schwefel03}. Note that infinite aperture limit is simply a Fourier transform of the incident field and we lose all the information about direction $\sin\chi$, consistent with our intuition with conjugate variables. It has to be emphasized that image data only probes the farfield, and does not contain the ``near-field" details we would see in a typical numerical solution, nor does it contain information about the internally reflected components of the cavity field (see Appendix~\ref{app_lenstransform} for further details). On the other hand, it contains (with some finite resolution) the same information as the Husimi distribution of the emitting components of the field and does allows us therefore to put forward a ray interpretation of the emission and the internal resonance.

\begin{figure}[t]
\centering
\includegraphics[width=0.9\linewidth]{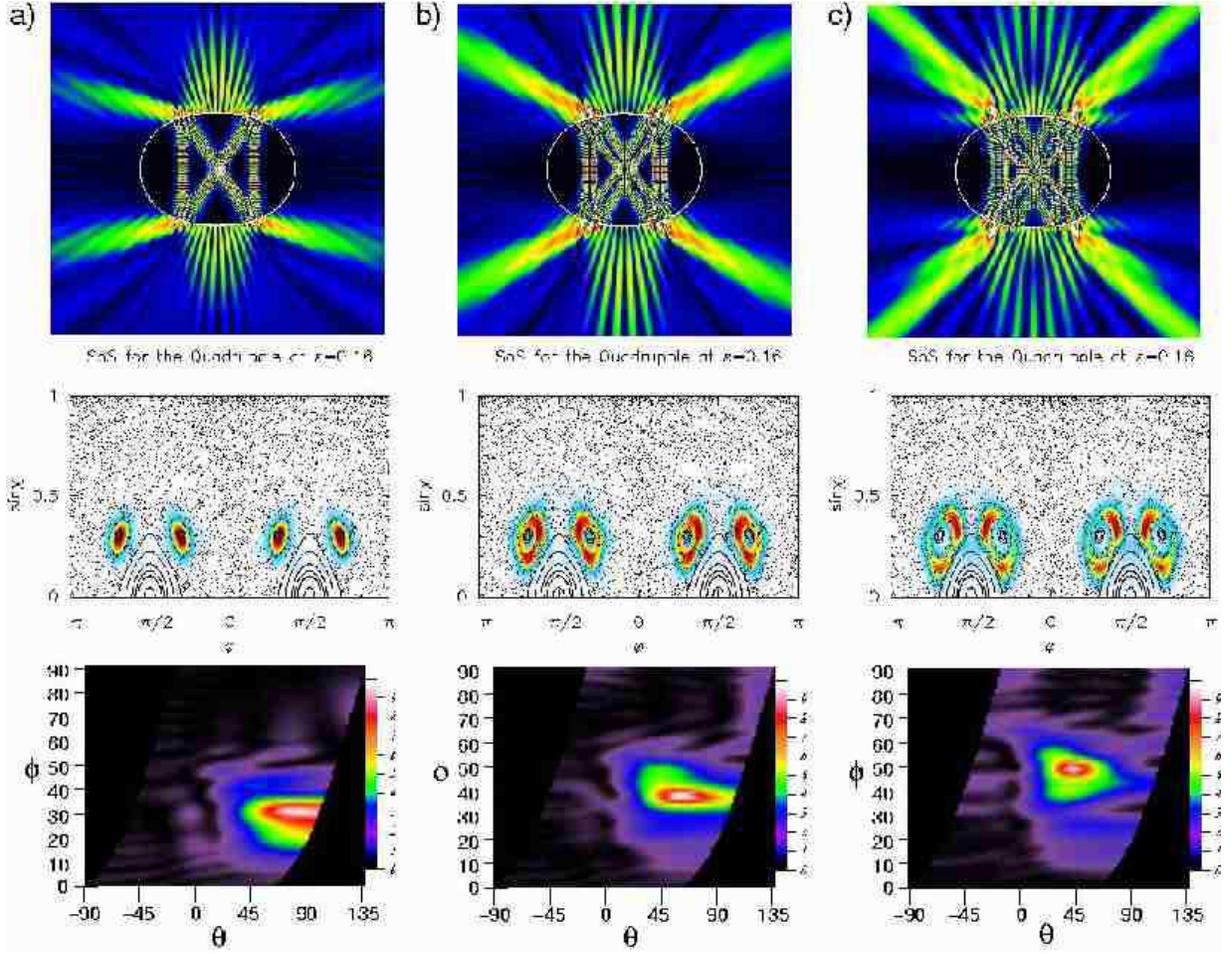}
\caption{A numerically calculated transverse series of modes based on the bow-tie periodic orbit. Such modes were observed in the experiments of ref.\cite{science}. The calculations are performed at $nkR\approx 128,\, n=3.3,\, \epsilon=0.16$. With these parameters the bowtie periodic orbit hits the boundary exactly at the critical angle. a) the fundamental mode, b) the first excited mode c) the second excited mode. Note the strong ``near-field'' fluctuations, particularly in the case of the second order mode in c). In the middle row we have their Poincar\'e-Husimi distributions. On the bottom we have for each of the modes a calculated boundary image, showing that one can clearly distinguish the three different modes from their boundary-image fields.}
\label{figthreebowties}
\end{figure}

The boundary-image field of a numerically obtained resonance can be calculated as described in Appendix~\ref{app_lenstransform}. In Fig.~\ref{figthreebowties} we plot the boundary-image fields of three ``bow-tie" modes of different transverse excitation number (based on a stable periodic orbit with the geometry of a bow-tie, see discussion below).

\section{Quasi-bound modes and classical phase space structures}
\label{sect_qbm}
In this section, we would like to illustrate the relation between the quasi-bound states and classical phase space structures using the results generated by the numerical algorithm described in section~(\ref{sect_rootsearch}) and Appendix~\ref{sect_numimp}. We note at the beginning that the real-space plots and the Husimi distributions are constructed using the {\em nonquantized} scattering eigenvectors, as we described in the previous section.

\begin{figure}[!hbt]
\centering
\includegraphics[width=0.7\linewidth,clip]{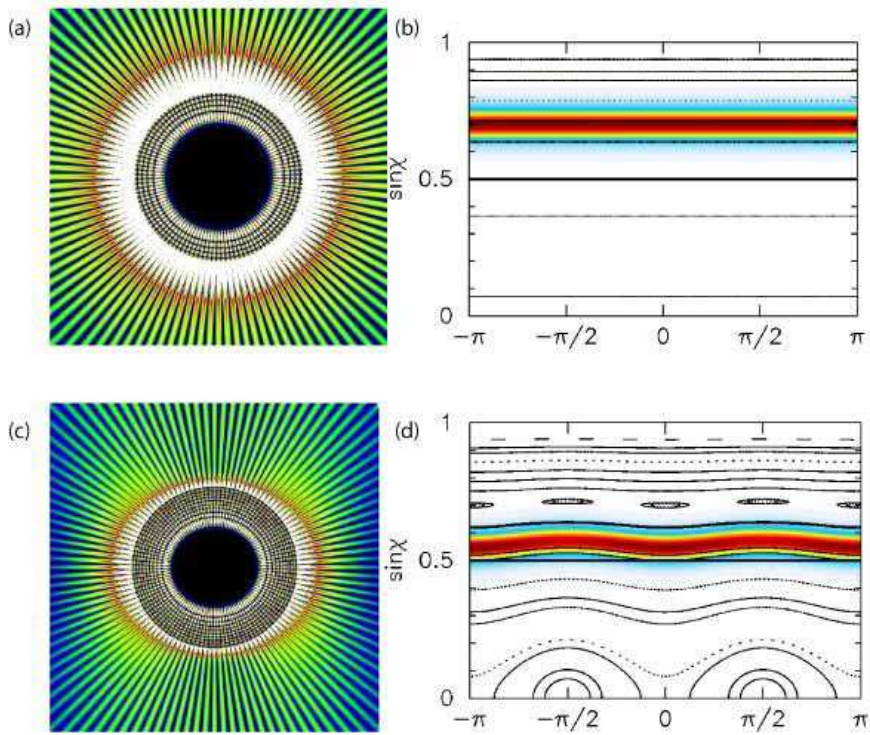}
\caption{Real-space false color plot of regular quasi-periodic solutions at (a) $\eps=0$, and (c) $\eps=0.03$. (b), (d) Husimi projections of states (a), (c). The solutions are obtained at $nkR=82$ and $n=2$.}
\label{figlazutkin}
\end{figure}

Consider first the near-integrable regime in the quadrupolar deformation. In Fig.~\ref{figlazutkin}(a), we plot a whispering gallery mode of the circle at $\eps=0$ and in Fig.~\ref{figlazutkin}(b) is plotted a state which emits from the highest curvature points $\phi=0,\pi$. Inspection of the respective Husimi distributions shows clearly that the first state is localized on an invariant curve $\sin\chi=const.$ and is (nearly) totally internally reflected, because the localization is at $\sin\chi>\sin\chi_c$.  Note that the second state is a deformed whispering gallery mode, localized on an invariant curve which is no longer a straight line in the SOS.

Large islands in the SOS (corresponding to stable periodic orbits) support multiple modes, which can be calculated approximately by the methods of Gaussian optics\cite{Tureci02}. We have already seen a typical sequence of such modes based on the bow-tie periodic orbit in Fig.~\ref{figthreebowties}. In general, it's possible to find such series of modes corresponding to any stable island of sizable extent in the SOS, with two characteristic spacings, the free spectral range $\Delta k = 2 \pi/L$ ($L$ is the length of the periodic orbit), and the transverse mode spacing, which depends on the eigenvalues of the stability matrix\cite{Tureci02}.

\begin{figure}[t]
\centering
\includegraphics[width=0.7\linewidth,clip]{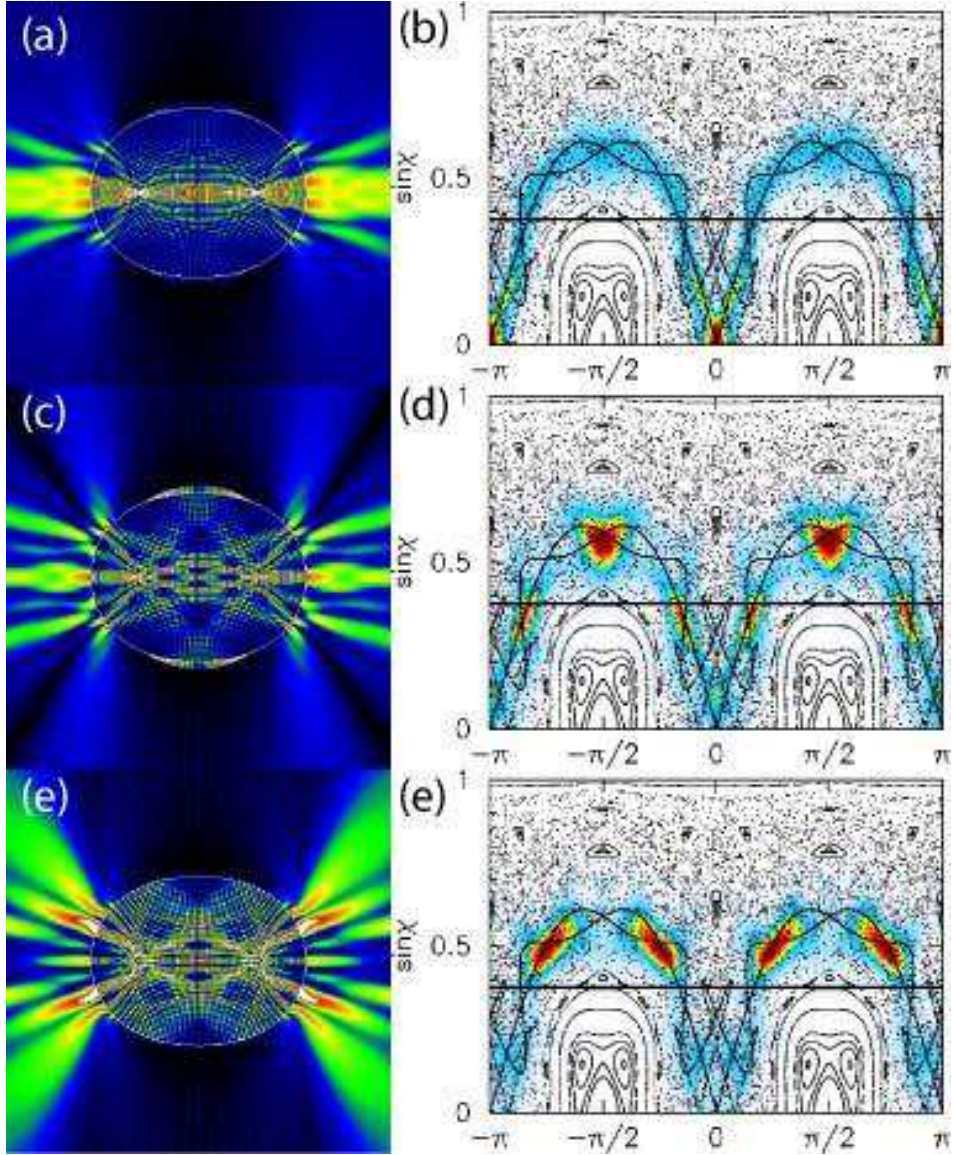}
\caption{Real-space plots and Husimi distributions of modes which are related to the unstable bouncing ball orbit (scars, see text). Superimposed on the SOS are the stable and unstable manifolds of this orbit which seem to play a role in localizing the solutions in certain regions of the chaotic component of phase space. (a), (b) represent a simple scar of the two-bounce orbit; (c), (d) and (e), (f) are states localized on intersections of the manifolds and appear more chaotic in real space. These solutions are found at $nkR=106$, $\eps=0.12$ and $n=2.65$.}
\label{figubbmodes}
\end{figure}

\begin{figure}[!hbt]
\centering
\includegraphics[width=0.7\linewidth,clip]{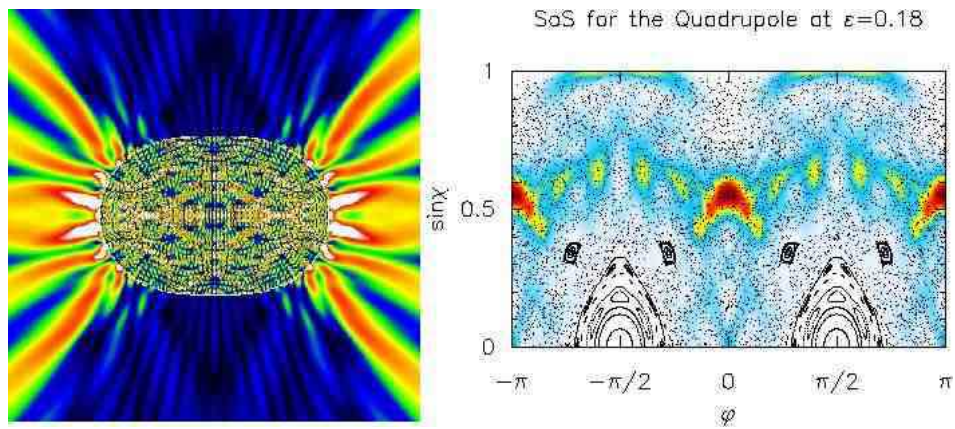}
\caption{Real-space false-color plot and Husimi distribution of a chaotic mode at a quadrupolar deformation of $\varepsilon=0.18$ and $n=2.65$, quantized at $kR=32.6638-0.06964i$.}
\label{figextremechaos}
\end{figure}

The importance of periodic orbits is not limited to stable orbits. There are an infinite number of unstable periodic orbits in the SOS. Short periodic orbits, especially those which are least unstable, can make their presence felt in the mode structure of the resonators, despite the fact that the methods of Gaussian optics fail for such modes\cite{Tureci02}. Such modes are referred to as ``scars" in the quantum chaos literature; they display an enhanced intensity along an unstable periodic orbit and have been widely studied\cite{heller84,kaplan99}. In the numerical studies of Refs.\cite{prosen93,frischat97}, evidence was found that modes can localize not only on the unstable fixed points but on their associated stable and unstable manifolds as well. In Fig.~\ref{figubbmodes}(a)-(f) we show a series of three states associated with the shortest unstable periodic orbit of the system, the unstable bouncing ball orbit. Note that the first mode represents the ``fundamental" mode which localizes on the fixed point itself (see Fig.~\ref{figubbmodes}(b)). The real-space plots Fig.~\ref{figubbmodes}(c) and Fig.~\ref{figubbmodes}(e) don't show any distinct structure. Their respective Husimi plots Fig.~\ref{figubbmodes}(d) and Fig.~\ref{figubbmodes}(f), however reveals that the modes localize on the heteroclinic intersections of the stable and unstable manifolds emanating from the unstable bouncing ball fixed points. A similar behavior was reported in Ref.\cite{frischat97} in the context of a quantum billiard.
Finally, at large deformations the spectrum contains chaotic modes which cannot easily be associated with any particular classical phase space structure. One such mode is plotted in Fig.~\ref{figextremechaos}. Note that the support of the mode is entirely in the chaotic portion of the SOS.  Recalling our arguments in section~(\ref{sect_torus_quant}) as to the failure of eikonal methods, it is instructive to note here the complexity of the wavefronts and the large portions of the resonator in which no series of parallel wavefronts is discernible.

\section{Conclusion}

We have considered a general class of optical resonators which are based on deformations of cylindrical dielectric resonators. Such resonators are being studied for applications in integrated optics and optoelectronics, and for their intrinsic interest as wave-chaotic systems. The basic physics of such resonators can be best understood by application of methods from classical and quantum chaos, applicable in the short-wavelength limit.

We have pointed out the characteristic global breakdown of conventional geometric optics approaches due to the transition to chaos in the associated ray dynamics. While real-space ray tracing methods quickly become powerless with increasing deformation (degree of chaos), the essential structures are uncovered effectively in the Poincar\'e surface of section. Information about important physical properties of deformed dielectric resonators and lasers, such as emission characteristics, internal modal distributions, spectra and lifetimes can be extracted from the types of ray motion in the equivalent refractive billiard system.  However no general analytic technique exists for calculating approximately all of the modes of a generic resonator of this type.

We present an efficient numerical method for the calculation of quasi-bound modes of dielectric cavities. The method we are proposing is a hybrid between a point-matching technique\cite{manenkov94} and scattering approach to quantization\cite{doron92,frischat97}. In contrast to existing methods\cite{manenkov94,noeckel_thesis,HentschelR02} which employ the external scattering matrix to extract the quasi-bound modes of a dielectric resonator, we consider the {\em internal scattering operator}\cite{doron92,DietzEPSU95,EckmannP95}. An important conceptual difference with respect to wave-function matching methods is that the internal scattering approach permits the identification  of a {\em discrete set} of internal scattering states at each value of $k$. This is realized by writing the matching conditions in the form of an off-shell eigenvalue problem instead of a linear inhomogeneous equation. Many of the physical properties of the modes within a given linewidth of the order of $1/nk_0R$ around $k_0$ are contained in the eigenvectors of the internal S-matrix at $k_0$, and can be extracted without quantizing the mode (i.e. without tuning $k$ to satisfy the boundary conditions). The quantized spectrum and the exact quasi-bound modes can be easily accessed by an extrapolation technique which requires only two diagonalizations of this S-matrix. In principle variants of this technique can be extended to very high wavevectors $k$ since it scales only  the perimeter of the resonator.

Finally, we present a version of the Husimi projection technique well-suited to dielectric resonators and show that the scattering eigenvectors as well as quasi-bound modes are structured by classical phase-space structures.

In conclusion, the methods and tools presented in this work provide a unified conceptual framework for treating dielectric resonators beyond the standard numerical approaches to such electromagnetic problems.

\begin{acknowledgments}
We would like to thank Evgenii Narimanov and Gregor Hackenbroich for their contributions to the early phase of the development of the numerical method.  We thank  Martin Gutzwiller for making us aware of Einstein's 1917 paper and for helpful discussions of its main point.  Special thanks to Nathan Rex and Richard Chang for providing us with experimental data to to compare with our numerical calculations. ADS would like to acknowledge support from NSF grant DMR-0084501 and PHJ lacknowledges support from the Swiss National Science Foundation. 
\end{acknowledgments}

\appendix
\section{Numerical Implementation Issues}
\label{sect_numimp}
Although the development in the text using the eigenvalue problem for the truncated S-matrix $\mcal{S}(k)$ is perfectly sound, its direct implementation is not very efficient.  The numerical problems already surface at the stage of computing the truncated scattering matrix $\mcal{S}(k)$ itself. The expression given in  Eq.~(\ref{openSmatform}) requires numerical inversion of 5 matrices. At a given $k$, as the number of channels $\Lambda$ of the interior matrices designated by index $1$ grows beyond $\Lambda_{sc}$, they become increasingly more singular.

As evident from our previous discussion, this ill-conditioning is caused by blindly including evanescent channels in the scattering problem. A quick solution is to truncate the matrices at the singularity boundary suggested by our ray interpretation and include only $\Lambda_{sc} = [\![nkR_{min}]\!]$ channels of $|m|$. But this turns out to produce some states which do not satisfy the boundary conditions well enough. As noted in\cite{doron92}, some of the evanescent channels $\Lambda_{ev}$ have to be kept, enough to be able to proceed with our numerical computation and provide the missing (evanescent) components of those states which require it. Thus, the properly truncated scattering matrix will have a size of $N_{trunc} = 2(\Lambda_{sc} + \Lambda_{ev}) + 1$.

The conditioning of $\mcal{S}(k)$ is highly sensitive to the choice of $\Lambda_{ev}$. Singular value decomposition can be employed to calculate the inverses and build up $\mcal{S}(k)$. However, one way to sidestep this issue in favor of a more robust method is to trade the eigenvalue problem Eq.~(\ref{evalproblem}) for a {\em generalized} eigenvalue problem. We rewrite
Eqs.~(\ref{Hcond1}-\ref{Hcond2}) in the form
\be
A\bra{\Upsilon} = \ex{i\varphi} B \bra{\Upsilon}
\ee
where the $2N_{trunc} \times 2N_{trunc}$ matrices are given by
\be
A = \mat{{\cal H}_1^+}{-{\cal H}_2^+}{{\cal DH}_1^+}{-\frac{1}{n}{\cal DH}_2^+}, \qquad B =
\mat{-{\cal H}_1^-}{0}{-{\cal DH}_1^-}{0}
\ee
and
\be
\bra{\Upsilon} = \vect{\bra{\alpha}}{\bra{\gamma}}
\ee
This way, the common null-space of the matrices $A$ and $B$ can be removed by existing powerful generalized eigenvalue solvers, such as {\tt ZGGEV} of {\tt LAPACK} library. As a byproduct we get both the inside and the outside vectors at one shot. This method turns out to be more stable than the one based on inversion, and yields good results with arbitrarily large $\Lambda_{ev}$. The numerical problems associated with the regions of evanescent behavior ($R_{min}<r<R_{max}$) remain, but are tractable for  $nkR\stackrel{<}{\sim}200-300$ (this range is larger for modes without evanescent components).

\section{Lens transform}
\label{app_lenstransform}
In the experimental imaging system, radiation emanating from the resonator is collected through an aperture and after passing through a lens, an image is recorded for a discrete number of angles in the farfield. The resonator is placed at the focal plane of the lens, so that the image is effectively formed at infinity.
\begin{figure}[!hbt]
\psfrag{psi1}{$\Psi(x)$}
\psfrag{psi2}{$\Psi^\prime(x)$}
\psfrag{psi3}{$\Psi^{\prime\prime}(u)$}
\centering
\includegraphics[width=0.5\linewidth]{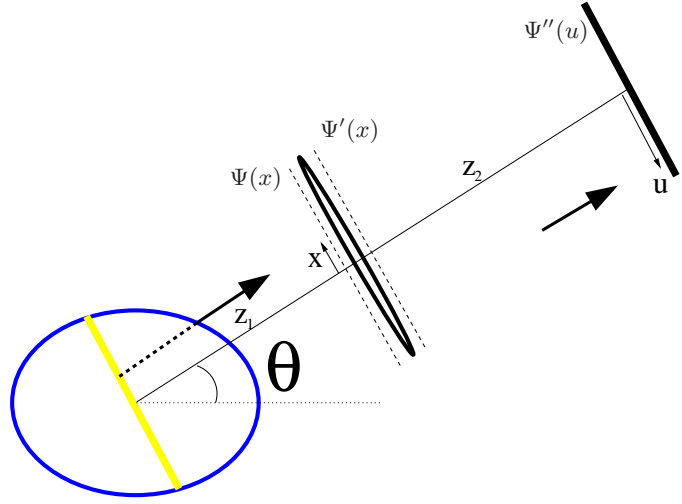}
\caption{Variables used in the lens transform to generate the boundary image field as measured experimentally.}
\label{figlenstransform}
\end{figure}
Just in front of the lens the field distribution is given by the resonance wavefunction $\Psi(x)$ which, at the observation (FF) angle $\theta$, can be expressed as:
\be
\Psi(x) \sim \sum_m \gamma_m \beshp{m}{k\sqrt{x^2+z_1^2}}
\mbox{e}^{im(\phi+\theta)}
\label{eqwfexp}
\ee
where $\phi=\tan^{-1}\frac{x}{z_1} \approx \frac{x}{z_1}$. The lens effectively adds a quadratic phase, so that the field immediately behind the lens is given by
\be
\Psi^{\prime}(x) = \Psi(x) P(x) \exp{\left[-i\frac{k}{2f}x^2\right]}
\ee
Here $P(x)$ is the {\em pupil function}, which takes care of the effect of the aperture, $f$ is the focal length of the lens and $x$ is the position on the lens. The field at the camera is given by propagating this field with the Fresnel propagator\cite{goodman_book}, which is well-justified as the lens-camera distance is much larger than the wavelength
\be
\Psi_{\theta}^{\prime\prime} (u) =
\frac{k}{iz_2}\int_{-\infty}^{\infty} dx \Psi^{\prime}(x)
\exp{[i\frac{k}{2z_2}(u-x)^2]}
\ee
Using the expression Eq.~(\ref{eqwfexp}) for the wavefunction
\bea
\Psi_{\theta}^{\prime\prime} (u) & = &
\frac{k}{iz_2}\mbox{e}^{(i\frac{k}{2z_2}(u-x)^2)}\sum_m \gamma_m \\
& & \times \int_{-\infty}^{\infty} dx P(x)
\beshp{m}{k\sqrt{k\sqrt{z_1^2+x^2}}}  \mbox{e}^{im(\theta +
\frac{x}{z_1})} \mbox{e}^{i\frac{k}{2}(\frac{1}{z_2}-\frac{1}{f})x^2}
e^{-i\frac{k}{z_2}ux}
\eea
Using the large argument asymptotic expansion of the Bessel functions:
\be
\beshp{m}{k\sqrt{x^2+z_1^2}} \sim \sqrt{\frac{2}{k(x^2+z_1^2)^{1/2}}}
\exp\left[ik\sqrt{x^2+z_1^2}-im\frac{\pi}{2}-i\frac{\pi}{4}\right]
\ee
Expanding the square root to ${\cal O}(\frac{x}{z_1})4$, i.e.\ to the same order as the Fresnel approximation and rearranging the terms,
\bea
\Psi_{\theta}^{\prime\prime} (u) & = &\frac{k}{iz_2}\mbox{e}^{i\frac{k}{2z_2}
u^2} \sum_m \gamma_m \frac{1}{kz_1}
\mbox{e}^{ikz_1+im\theta-im\frac{\pi}{2}-i\frac{\pi}{4}} \\
& & \times \int_{-\infty}^{\infty} dx P(x)
\mbox{e}^{im\frac{x}{z_1}-\frac{k}{z_2}ux}
\mbox{e}^{i\frac{k}{2}(\frac{1}{z_1}+\frac{1}{z_2}-\frac{1}{f})x^2}
\eea
The second exponent in the integral is exactly the lens law, so it vanishes. Setting $\frac{m}{k}=R_o\sin\chi_m$, the intensity recorded at the pixel $u$ of the camera at the farfield angle $\theta$ can asymptotically be written as
\be
\left|\Psi_{\theta}^{\prime\prime} (u)\right|^2 \sim \left| \sum_m
\gamma_m \beshp{m}{kz_1} \mbox{e}^{im\theta} \int_{-\infty}^{\infty}
dx P(x) \exp{\left[i\frac{k}{z_2}(MR_o\sin\chi_m-u)x\right]} \right|^2
\ee
where $M=z_2/z_1$ is the magnification of the lens. For a simple aperture, $P(x)$ is just a rectangle function, so that the integral can be performed exactly to yield
\be
\left|\Psi_{\theta}^{\prime\prime} (u)\right|^2 \sim \left| A \sum_m
\gamma_m H_m^{+} (kz_1) \mbox{e}^{im\theta} \mbox{sinc}
\left[\frac{1}{\Delta} (\sin\chi_m - \frac{u}{M})\right]\right|^2
\label{eqif}
\ee
where $\Delta=\frac{2z_1}{AkR_o}$ and $\mbox{sinc}(x) = \sin x/x$. Note that in the short-wavelength limit $\Delta \rightarrow 0$ and $\frac{1}{\pi\Delta}\mbox{sinc}(\frac{x}{\Delta}) \rightarrow \delta(x)$. This expression allows us to make predictions based on short-wavelength limit and geometric ray optics, which includes effects of diffraction as well.  For instance, for a circular cylindrical resonator, the resonances are composed of a single angular momentum component $m$ (and its degenerate partner $-m$). In that case, according to the expression Eq.~(\ref{eqif}),
\be
\left|\Psi_{\theta}^{\prime\prime} (u)\right|^2 \propto
\left|\delta(\sin\chi_m - \frac{u}{M}) \mbox{e}^{im\theta} \pm
\delta(\sin\chi_m + \frac{u}{M}) \mbox{e}^{-im\theta}\right|^2
\ee
Note that the imagefield contains only information captured from the farfield distribution. The actual details of the resonance in the ``nearfield" can be quiet different, due to evanescent contributions close to critical incidence. For instance, the points of brightest emission inferred from the imagefield might be shifted due to an ``optical mirage''-like effect (see Fig.~\ref{figthreebowties}c)). The mirage is formed not because of a continuously varying index of refraction but a discontinuous interface.

The imagefield has an interesting connection to the (SOS projected) Husimi distribution. The Husimi distribution of the field projected onto the SOS at a distance $R \rightarrow \infty$ is given by
\be
H_{\Psi} (\theta,p_{\theta}) = \left| \sum_m \gamma_m \beshp{m}{kR}
\mbox{e}^{im\theta} \mbox{e}^{-\frac{1}{2} \eta ^2 (m -
p_{\theta})^2} \right|^2
\ee
Comparing with Eq.~(\ref{eqif}), we see that the two functions are contain almost the same information. In fact, by choosing an aperture which has a gaussian transmittance $P(x)$, one would obtain exactly the same form as Eq.~(\ref{eqif}). Note that the freedom of smoothing to obtain various phase space distributions which represent the same physical system gains here a physical meaning, namely it translates to the choice of optical apparatus (lens, aperture etc.) to observe the resonator. This connection was used in reference\cite{schwefel03} to reconstruct from the boundary-image field a Husimi function for the emitted radiation which was found to agree well with ray escape simulations.


\end{document}